\documentclass[aps,prl,a4paper,twocolumn]{revtex4-2}

\usepackage{bm}
\usepackage{amsmath}
\usepackage{amsfonts}
\usepackage{amssymb}
\usepackage{color}
\usepackage{graphicx}
\usepackage{changes}
\usepackage[final]{pdfpages}
\usepackage{pgffor}

\makeatletter
\AtBeginDocument{\let\LS@rot\@undefined}
\makeatother

\begin{document}

\title{Bath-induced spin inertia}


\author{Mario Gaspar Quarenta$^1$}
\author{Mithuss Tharmalingam$^{1,2}$}
\author{Tim Ludwig$^{1,3}$}
\author{H.~Y.~Yuan$^{1,4}$}
\author{Lukasz Karwacki$^{1,5}$}
\author{Robin C. Verstraten$^1$}
\author{Rembert Duine$^{1,6}$}

\affiliation{$^{1}$Institute for Theoretical Physics, Utrecht University, \\
Princetonplein 5, 3584 CC Utrecht, The Netherlands;}

\affiliation{$^{2}$Department of Engineering Sciences, Universitetet i Agder, \\
Postboks 422, 4604 Kristiansand, Norway;}

\affiliation{$^{3}$Department of Philosophy, Institute of Technology Futures,\\
Karlsruhe Institute of Technology, Douglasstraße 24, 76133 Karlsruhe, Germany;}

\affiliation{$^{4}$Department of Quantum Nanoscience, Kavli Institute of Nanoscience, \\
Delft University of Technology, Lorentzweg 1, 2628 CJ Delft, The Netherlands;}

\affiliation{$^{5}$Institute of Molecular Physics, Polish Academy of Sciences,\\
ul. M. Smoluchowskiego 17, 60-179 Pozna\'{n}, Poland;}

\affiliation{$^{6}$Department of Applied Physics, Eindhoven University of
Technology, \\
P.O. Box 513, 5600 MB Eindhoven, The Netherlands}

\date{\today}

\begin{abstract}
Spin dynamics is usually described as massless or, more precisely, as free of inertia. Recent experiments, however, found direct evidence for inertial spin dynamics. In turn, it is necessary to rethink the basics of spin dynamics. Focusing on a macrospin in an environment (bath), we show that the spin-to-bath coupling gives rise to spin inertia. This bath-induced spin inertia appears universally from all the high-frequency bath modes. We expect our results to provide new insights into recent experiments on spin inertia. Moreover, they indicate that any channel for spin dissipation should also be accompanied by a term accounting for bath-induced spin inertia. As an illustrative example, we consider phonon-bath-induced spin inertia in a YIG/GGG stack.
\end{abstract}

\maketitle

\textit{Introduction.}---Spin dynamics is usually considered to be free of inertia and, in turn, it is described by first-order differential equations; for example, the Bloch equation \cite{bloch1946} for small quantum-mechanical spins or the Landau-Lifshitz-Gilbert (LLG) equation \cite{lifshitz1935, landau1992, gilbert1956, gilbert2004} for larger quasi-classical spins. Based on the broad success of those descriptions, one might be tempted to conclude that spin dynamics is free of any inertia, as spin inertia would be described by second-order time derivatives. Recent experiments, however, find direct evidence for spin inertia close to $\mathrm{THz}$-frequencies \cite{neeraj2021, unikandanunni2022}; see also \cite{li2015}. First and foremost, the discovery of spin inertia is of immense interest for our fundamental understanding of spin dynamics. For example, spin inertia leads to a redshift in the ferromagnetic resonance peak \cite{cherkasskii2022theory, cherkasskii2020nutation} and, more interestingly, gives rise to nutation spin waves \cite{mondal2022inertial, cherkasskii2021dispersion, titov2022nutation}. In addition, spin inertia might also prove relevant in applications as the related nutation dynamics takes place on short time scales \cite{neeraj2021, unikandanunni2022, li2015, fahnle2019, mondal2022inertial}; for example, spin inertia might allow for faster and more efficient spin switching similar to antiferromagnets \cite{kimel2009}. For a recent review on inertial effects, see reference \cite{mondal2023inertial}. 

Spin inertia and spin nutation have already been derived for several systems with various approaches: from an adiabatic expansion of a dissipation kernel \cite{suhl1998theory}; for a spin in a superconducting Josephson junction \cite{zhu2004novel}; from mesoscopic nonequilibrium thermodynamics \cite{ciornei2011}; with a classical Lagrangian approach \cite{wegrowe2012} analogous to the derivation of Gilbert damping \cite{gilbert2004}; for the spins coupled to an electron bath \cite{fahnle2011, bhattacharjee2012, kikuchi2015, hurst2020}; from higher-order relativistic terms \cite{mondal2017, mondal2018}; for environments with a Lorentzian bath spectral density \cite{anders2022quantum}; and by modeling a magnetization in terms of a current-carrying loop \cite{titov2021}. Despite the many crucial insights provided by these approaches, a complete understanding of spin inertia is arguably still missing \cite{unikandanunni2022}. Here, in the quest to contribute to a more general and unified understanding of spin inertia, we show that spin inertia arises universally from the interaction with an environment.

\begin{figure}[b]
\begin{center}
\includegraphics[width=0.35\textwidth]{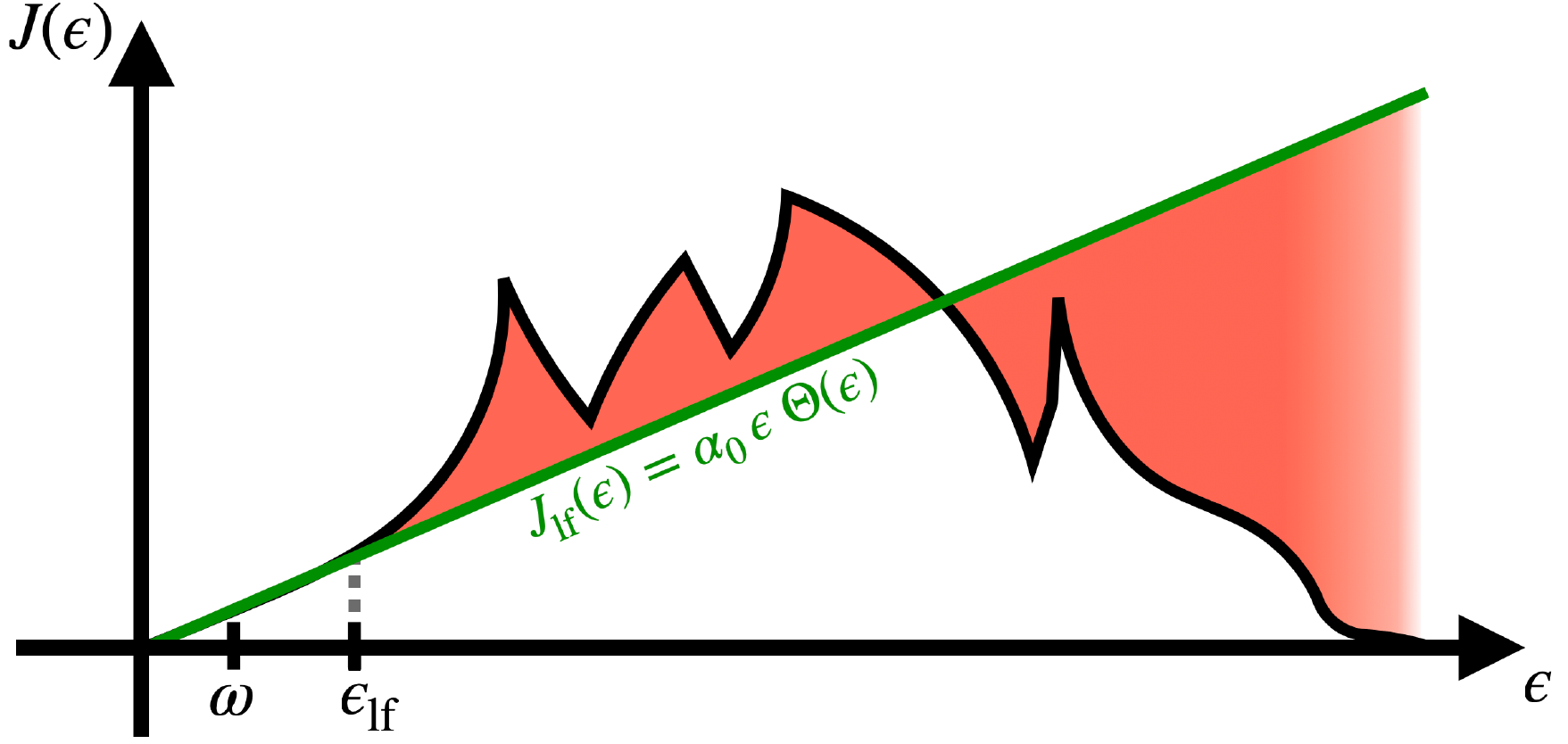}
\end{center}
\caption{The bath spectral density $J(\epsilon)$ contains the information about how the environment (bath) affects the spin dynamics. We separate $J(\epsilon)$ into a low-frequency approximation $J_\mathrm{lf}(\epsilon)$ (green solid line) and the remaining high-frequency part $J(\epsilon) - J_\mathrm{lf}(\epsilon)$ (red shaded area). This separation sets an energy scale, $\epsilon_\mathrm{lf}$, up to which $J_\mathrm{lf}(\epsilon)$ is a good approximation to $J(\epsilon)$. Assuming the typical frequencies of the spin dynamics to be small, $\omega \ll \epsilon_\mathrm{lf}$, the low-frequency bath modes give rise to damping, whereas all the high-frequency bath modes give rise to spin inertia. While our results are general, for the figure we assumed $J_\mathrm{lf}(\epsilon)$ to be linear, which leads to Gilbert damping.} \label{fig:spectral}
\end{figure}

In this letter, we consider a macrospin that is exposed to an effective magnetic field and coupled to its environment. The interaction with the effective magnetic field is described by the Zeeman energy. Using the Caldeira-Leggett approach \cite{caldeiraandleggett1983, caldeiraandleggett1983a}, analogous to \cite{verstraten2022}, we model the environment as a bath of harmonic oscillators and, for simplicity, assume a linear coupling between the macrospin and the bath modes. In the quasi-classical limit, after integrating out the bath modes, we find an inertial LLG equation. More explicitly, we show that the environmental degrees of freedom (bath modes) affect spin dynamics in two ways: the low-frequency bath modes give rise to Gilbert damping (or fractional Gilbert damping if the bath is non-ohmic), while all the high-frequency bath modes give rise to spin inertia (independent of the bath details); see Fig. \ref{fig:spectral}.

Finally, to show that the Caldeira-Leggett approach with linear coupling is not just a toy model but of real use, we discuss its application to a macrospin that is coupled to a phonon bath and we derive the phonon-bath-induced spin inertia for a magnetic plane (YIG) that is sandwiched between two nonmagnetic insulators (GGG).

\textit{Model Hamiltonian.}---We consider a macrospin that is coupled to its environment and exposed to an (effective) magnetic field. The corresponding Hamiltonian can be separated into three parts, $\hat H = \hat H_s + \hat H_c + \hat H_b$, where the system Hamiltonian $\hat H_s$ describes the macrospin in the magnetic field, the bath Hamiltonian $\hat H_b$ describes the environmental degrees of freedom, and the coupling Hamiltonian $\hat H_c$ describes the coupling between the macrospin (system) and its environment (bath). Explicitly, the macrospin in the magnetic field is described by the Zeeman energy with the Hamiltonian $\hat H_s = - \mathbf B \cdot \mathbf{\hat S}$, where $\mathbf{\hat S}= (\hat S_x, \hat S_y, \hat S_z)$ is the macrospin operator with standard spin operators $\hat S_x$, $\hat S_y$, $\hat S_z$ and $\mathbf B$ denotes the (effective) magnetic field that may dependent on time and can point in any direction. Using the Caldeira-Leggett approach \cite{caldeiraandleggett1983, caldeiraandleggett1983a}, we model the environment as a bath of harmonic oscillators; in turn, the bath Hamiltonian is given by $\hat H_b = \sum_n (\mathbf{\hat p}^2_n/2m_n + m_n \omega_n^2 \mathbf{\hat q}^2_n/2 )$, where $m_n$, $\omega_n$, $\mathbf{\hat q}_n$, $\mathbf{\hat p}_n$ are mass, eigenfrequency, position operator, and momentum operator of the $n$-th bath oscillator. For simplicity, we assume a linear system-to-bath coupling, which is described by the coupling Hamiltonian $\hat H_c = - \sum_n \gamma_n\, \mathbf{\hat q}_n \cdot \mathbf{\hat S}$, where $\gamma_n$ is the coupling coefficient describing the coupling strength between the macrospin and the $n$-th bath oscillator respectively. The rationale behind the linear coupling is that, close to the ground state of the full system (macrospin $+$ environment), a Taylor expansion is likely to lead to a linear coupling as the leading order term. Thinking in more physical terms, the linear coupling might, for example, be an \textit{sd}-like coupling between the macrospin and electron-hole pairs in metallic magnets. However, as we will show below, even in the case of magnetoelastic coupling, which is quadratic in the spin variables, the linear-coupling approach is still useful to understand the inertial macrospin dynamics close to the magnetic ground state.

\textit{Effective action for spin dynamics.}---Starting from the model, we use the Keldysh formalism in its path-integral version \cite{kamenevandlevchenko2009, altlandandsimons2010, kamenev2011} with spin coherent states $|g\rangle$, as in reference \cite{altlandandsimons2010}, to derive an action for the combined dynamics of the macrospin and the bath oscillators. Because we model the environment as a bath of harmonic oscillators and assume linear coupling, the action is (at most) quadratic in $\mathbf{\hat p}_n$ and $\mathbf{\hat q}_n$. In turn, we can integrate out the bath modes by performing Gaussian path integrals; first in $\mathbf p_n$, then in $\mathbf q_n$. As result, we obtain the Keldysh partition function $\mathcal{Z}  = \int Dg\, e^{i \mathcal{S}}$ with the effective action for the macrospin dynamics
\begin{equation}
\mathcal S\! =\! \oint_K\!\!\! dt \left[ -i \langle \dot g | g \rangle\! +\!
\mathbf B\! \cdot\! \mathbf S \right] + \frac{1}{2}\! \oint_K\!\!\! dt\!
\oint_K\!\!\! dt'\, \mathbf S(t)\, \alpha(t-t')\, \mathbf S(t')\,
,\label{eq:action}
\end{equation}
where $|\dot g\rangle = \partial_t | g \rangle$ and $\mathbf S = \langle g | \mathbf{\hat S} | g \rangle$. Information about the coupling to the bath is contained in the kernel function $\alpha(t-t')$ that, when represented in Keldysh space, is a matrix containing as elements the retarded and advanced parts $\alpha^{R/A}(\omega)$ and a Keldysh part $\alpha^{K}(\omega)$. The noiseless effects (damping and inertia) are described by the retarded and advanced parts $\alpha^{R/A}(\omega) = - \sum_n \gamma_n^2/\lbrace  m_n\, [(\omega \pm i \eta)^2 - \omega_n^2]\rbrace$, where $\eta$ is an infinitesimal level broadening or decay rate. The information about fluctuations (noise) is contained in the Keldysh part $\alpha^{K}(\omega) = \coth(\omega/2k_B T)\ [ \alpha^{R}(\omega) - \alpha^{A}(\omega)]$ with Boltzmann constant $k_B$ and bath temperature $T$, which is simply the fluctuation-dissipation theorem, because we assume the bath to be in (local) equilibrium \cite{kamenev2011}.

To be more explicit, as in reference \cite{altlandandsimons2010}, we use the Euler-angle representation of spin coherent states, $|g\rangle = \exp(- i \phi \hat S_z)\, \exp(- i \theta \hat S_y)\, \exp(- i \psi \hat S_z)\, \left| \Uparrow \right\rangle$, where $\left| \Uparrow \right\rangle$ is the eigenstate to $\hat S_z$ with the maximal eigenvalue $S$; that is, $\hat S_z\, \left|\Uparrow \right\rangle = S\, \left| \Uparrow \right\rangle$. This representation is convenient, as the macrospin takes the intuitive form of a vector in spherical coordinates $\mathbf S = S\, (\sin \theta\cos \phi,\, \sin \theta \sin \phi,\, \cos \theta)$, where $S$ is its length and $\theta$ and $\phi$ describe its orientation. The Berry-phase term becomes $-i \langle \dot g | g \rangle = S\, (\dot \psi + \dot \phi \cos \theta)$, where the angle $\psi$ is a gauge freedom \cite{altlandandsimons2010}. Properly fixing this gauge on the Keldysh contour is nontrivial and can be crucial \cite{shnirman2015, ludwig2019}. Here, however, a simple choice of $\dot \psi = - \dot \phi$ on the Keldysh contour will be just fine; for a detailed discussion, see the Supplemental Material.

\textit{Generalized Landau-Lifshitz-Gilbert equation.}---From the effective action, we derive an effective quasi-classical equation of motion for the macrospin along the following lines: first, to retain information about fluctuations, we use the Schmid trick \cite{schmid1982, eckern1990, altlandandsimons2010, kamenev2011, weiss2012} and ``decouple'' the part of the action that is quadratic in quantum components by a Hubbard-Stratonovich transformation; then, we vary the action with respect to the quantum components $\theta_q$ and $\phi_q$; finally, the resulting quasi-classical equations of motion for $\theta_c$ and $\phi_c$ can be recast into a single vectorial equation of motion \footnote{For notational simplicity, since no quantum components remain, we drop the index $c$ for classical from now on.}. As result, we obtain a generalized LLG equation
\begin{equation}
\mathbf{\dot S} = \mathbf{S} \times (\mathbf{B} + \boldsymbol{\xi}) + \mathbf{S} \times 
\int_{-\infty}^{\infty}\!\!\! dt'\ \tilde \alpha (t-t')\, \mathbf{S}(t')\ , \label{eq:gLLG}
\end{equation}
where $\boldsymbol{\xi}$ is a fluctuating field with $\langle \xi_m(t)\rangle=0$ and $\langle \xi_m(t) \xi_{m'}(t') \rangle = -(i/2)\, \delta_{mm'}\, \alpha^K(t-t')$; the indices $m, m'\in \lbrace x,y,z\rbrace$ denote Cartesian components corresponding to $\mathbf S = (S_x, S_y, S_z)$.  Furthermore, we defined $\tilde \alpha (\omega) = \alpha^R(\omega) - \alpha^R(0)$, where the $\omega = 0$ part is subtracted for regularization \footnote{Note that we have used $\alpha^A(-\omega) = \alpha^R(\omega)$. Furthermore, note that an $\omega =0$ contribution as $\alpha^R(0)$ could renormalize the effective magnetic field. In the present case, however, the $\omega =0$ contribution is irrelevant for the quasi-classical dynamics, as it would lead to a term proportional to $\mathbf S \times \mathbf S$, which vanishes identically.}. More explicitly, the kernel function is given by
\begin{equation}
\tilde \alpha (\omega) = - \frac{2}{\pi} \int_{-\infty}^{\infty}\! d\epsilon\,
\frac{\omega^2 J(\epsilon)}{[(\omega + i \eta)^2 - \epsilon^2]\, \epsilon}\ ,
\label{eq:kernel}
\end{equation}
where $J(\epsilon) = \pi \sum_n (\gamma_n^2/ 2m_n \omega_n)\, \delta(\epsilon - \omega_n)$ is the bath spectral density that contains two pieces of information: in the delta function $\delta(\epsilon - \omega_n)$, it contains the information at which energies the bath modes can be found; in the pre-factor $(\gamma_n^2/2 m_n \omega_n)$ it contains the information of how strongly the macrospin couples to the bath modes.

\textit{Separating low- and high-frequency bath modes.}—
In an environment with many bath modes that are close in energy, the bath spectral density $J(\epsilon)$ will be a continuous function; see Fig. \ref{fig:spectral}. For a general bath spectral density, it is hard to proceed analytically. However, inspired by reference \cite{weiss2012} but without introducing a cutoff, we make analytical progress by separating the contributions of low-frequency and high-frequency bath modes. Explicitly, we rewrite $J(\epsilon) = J_\mathrm{lf}(\epsilon) + [J(\epsilon) - J_\mathrm{lf}(\epsilon)]$, where $J_\mathrm{lf}(\epsilon)$ is a low-frequency approximation of $J(\epsilon)$ and we refer to $J(\epsilon) - J_\mathrm{lf}(\epsilon)$ simply as high-frequency contribution, even though non-low-frequency contribution might be a more accurate name. Accordingly, we split the kernel function $\tilde \alpha(\omega) = \tilde \alpha_\mathrm{lf}(\omega) + \tilde \alpha_\mathrm{hf}(\omega)$ into a low-frequency contribution $\tilde \alpha_\mathrm{lf}(\omega)$ and the high-frequency contribution $\tilde \alpha_\mathrm{hf}(\omega)$. Explicitly, the low-frequency contribution is given by
\begin{equation}
\tilde \alpha_\mathrm{lf} (\omega) = - \frac{2}{\pi} \int_{-\infty}^{\infty}\!
d\epsilon\, \frac{\omega^2 J_\mathrm{lf}(\epsilon)}{[(\omega + i \eta)^2 -
\epsilon^2]\, \epsilon}\ ,  \label{eq:lf-kernel}
\end{equation}
while the high-frequency contribution is given by
\begin{equation}
\tilde \alpha_\mathrm{hf} (\omega) = - \frac{2}{\pi} \int_{-\infty}^{\infty}\!
d\epsilon\, \frac{\omega^2 [J(\epsilon) - J_\mathrm{lf}(\epsilon)]}{[(\omega +
i \eta)^2 - \epsilon^2]\, \epsilon}\ . \label{eq:hf-kernel}
\end{equation}

Now, the key point to realize is that we are able to determine the high-frequency contribution under a quite general assumption: we assume that typical frequency of the spin dynamics $\omega$ is much smaller than $\epsilon_\mathrm{lf}$, which is the energy up to which $J_\mathrm{lf}(\epsilon) \approx J(\epsilon)$; see Fig. \ref{fig:spectral}. Under this assumption, we can disregard the $\omega$ dependence in the denominator of Eq. \eqref{eq:hf-kernel}. The reason is that $\epsilon < \epsilon_\mathrm{lf}$ the numerator vanishes, $J(\epsilon) - J_\mathrm{lf}(\epsilon)\approx 0$, while for $\epsilon > \epsilon_\mathrm{lf}$ the denominator can be approximated, $(\omega + i \eta)^2 - \epsilon^2 \approx - \epsilon^2$. In turn, we can approximate the high-frequency contribution to the kernel function, 
\begin{equation}
\tilde \alpha_\mathrm{hf} (\omega) \approx I\,\omega^2\ ,
\label{eq:hf-kernel-approx}
\end{equation}
and we arrive at our central result: the bath-induced spin inertia
\begin{equation}
I = \frac{2}{\pi} \int_{-\infty}^{\infty}\! d\epsilon\ \frac{J(\epsilon) -
J_\mathrm{lf}(\epsilon)}{\epsilon^3}\ . \label{eq:inertia}
\end{equation}
So, if the bath spectral density $J(\epsilon)$ is known, it can be used to determine its low-frequency approximation $J_\mathrm{lf}(\epsilon)$ and, in turn, to find the spin inertia $I$ by straightforward (potentially numerical) integration. Note that the sign of the spin inertia is not fixed; depending on the form of $J(\epsilon)$, spin inertia can be positive or negative.

\textit{Inertial Landau-Lifshitz-Gilbert equation.}—
To recover the inertial LLG equation used in experimental analysis \cite{neeraj2021, unikandanunni2022}, we assume the bath spectral density to be approximately linear at low frequencies. This assumption allows us to use an Ohmic bath, which is a bath with linear bath spectral density, for the low-frequency approximation $J_\mathrm{lf}(\epsilon)$; for non-Ohmic low-frequency approximations, see reference \cite{verstraten2022}. Explicitly, we assume $J_\mathrm{lf}(\epsilon) = \alpha_0 \epsilon\, \Theta(\epsilon)$, where $\Theta(\epsilon)$ is the Heaviside $\Theta$-function and $\alpha_0$ is some constant that will turn out to be the Gilbert-damping coefficient. We then find $\tilde \alpha_\mathrm{lf} (\omega) = i \alpha_0 \omega$ for the low-frequency kernel function and $I = (2/\pi) \int_{0}^{\infty}\! d\epsilon\ [J(\epsilon) - \alpha_0 \epsilon]/\epsilon^3$ for the bath-induced spin inertia, where we used that bath spectral densities vanish for negative energies.

It is now straightforward to derive the inertial LLG equation. First, transforming the low- and high-frequency parts of the kernel function back into time space, we find $\tilde \alpha_\mathrm{lf} (t-t') = - \alpha_0\, \delta' (t-t')$ and $\tilde \alpha_\mathrm{hf}(t-t') = - I\, \delta''(t-t')$, where $\delta'(t)$ and $\delta''(t)$ are respectively the first and second derivative of the Dirac $\delta$-function. Then, inserting them back into the generalized LLG equation
\eqref{eq:gLLG}, we obtain the inertial LLG equation
\begin{equation}
\mathbf{\dot S} = \mathbf S \times \left[ \mathbf B + \boldsymbol{\xi} - \alpha_0\, \mathbf{\dot S} - I\, \mathbf{\ddot S}
 \right]\ , \label{eq:iLLG}
\end{equation}
where $\alpha_0$ is the Gilbert-damping coefficient, $I$ is the spin inertia, and $\boldsymbol \xi $ is a fluctuating field with $\langle \xi_m(t)\rangle=0$ and $\langle \xi_m(t) \xi_{m'}(t') \rangle \approx 2 k_B T\alpha_0\, \delta_{mm'}\, \delta(t-t')$ for which we assumed the temperature to be large \footnote{At low temperatures, the correlation function is given by $\langle \xi_m \xi_{m'} \rangle (\omega) = \alpha_0\, \omega \coth (\omega/2k_B T)\, \delta_{mm'}$, where $\omega$ is the frequency corresponding to $t-t'$. At high temperatures, $2k_B T \gg \omega$, the correlation function simplifies to $\langle \xi_m \xi_{m'} \rangle (\omega) = 2 \alpha_0\, k_B T\, \delta_{mm'}$.}. Note that only Gilbert damping—but not spin inertia—contributes to fluctuations. On the formal level, spin inertia does not contribute to fluctuations, as even-frequency contributions of $\alpha^{R}(\omega)$ and $\alpha^A(\omega)$ cancel out in $\alpha^K(\omega)$. In more physical terms, spin inertia is not dissipative and, based on the general idea of the fluctuation-dissipation theorem \cite{vankampen2011, weiss2012}, should therefore also not contribute to fluctuations.

The Gilbert-damping coefficient $\alpha_0$ and the spin inertia $I$ depend on the bath spectral density $J(\epsilon)$; see Eqs. \eqref{eq:lf-kernel} and \eqref{eq:hf-kernel} respectively. Next, to illustrate how our general results can be applied, we consider a macrospin that is coupled to a phonon bath.

\textit{A macrospin coupled to a phonon bath.}—
The coupling between spins and phonons can be described by magnetoelastic coupling, as in reference \cite{ruckriegelandkopietz2015}, where they derived a generalized LLG equation analogous to Eq. \eqref{eq:gLLG} but for a spin lattice. So, to identify the bath-induced spin inertia for phonon baths, we do not need to rederive the generalized LLG equation. Instead, we start from their generalized LLG equation \cite{ruckriegelandkopietz2015}, recast it into the form of our Eq. \eqref{eq:gLLG}, identify the bath spectral density $J(\epsilon)$, and finally determine the spin inertia from Eq. \eqref{eq:inertia}.

Note that magnetoelastic coupling is derived in the continuum limit of long-wavelength phonons. So, strictly speaking, one would have to rederive the spin-phonon coupling to include short-wavelength (high-frequency) phonon modes, which are responsible for spin inertia. Nevertheless, we believe that the following illustrative example based on magnetoelastic coupling provides valuable insights into the physics of spin inertia and paves the way for more detailed derivations from microscopic models.

Magnetoelastic coupling is quadratic in the spin variables \cite{ruckriegelandkopietz2015}. Close to the ground state, however, we are allowed to linearize the spin dynamics and with it the magnetoelastic coupling. Combining this linearization with a macrospin approximation, we relate the phonon bath \cite{ruckriegelandkopietz2015} to the Caldeira-Leggett approach above; for details, see Supplemental Material. Explicitly, we find that the kernel function \eqref{eq:kernel} becomes a tensor $\tilde \alpha_{mm'}(\omega)$, as also the bath spectral density takes a tensorial form \footnote{Here, the indices $m,m'$ are defined as above after \eqref{eq:iLLG}.},
\begin{align}
J_{m m'}(\epsilon)=&\ \pi \sum_i \sum_{\mathbf k\lambda} e^{i \mathbf k \mathbf
R_i} \frac{B_{m z} B_{m' z}}{2M N S_1 S\,\omega_{\mathbf k
\lambda}}\nonumber \\ & \times (\mathbf k_{m z} \cdot \mathbf e_{\mathbf
k\lambda})\,(\mathbf k_{m' z} \cdot \mathbf e_{-\mathbf k\lambda})\, \delta
(\epsilon - \omega_{\mathbf k \lambda})\ , \label{eq:J-ph}
\end{align}
where $S_1$ is the effective spin length on an individual lattice site with spin, $B_{m \tilde m}$ and $B_{m' \tilde m}$ are the magnetoelastic coupling coefficients (here $\tilde m=z$, as we chose the $z$-direction for the ground state), the scalars $N$ and $M$ are respectively the number and effective (oscillating) mass of lattice sites, the function $\omega_{\mathbf k\lambda}$ is the phonon dispersion relation with the phonon momentum $\mathbf k$, the vector $\mathbf e_{\mathbf k \lambda}$ is the polarization vector and $\lambda$ is the index denoting longitudinal and transversal polarization, the vector $\mathbf k_{mm'}$ is defined by $\mathbf k_{mm'} = k_m \mathbf e_{m'} + k_{m'} \mathbf e_m$ with Cartesian unit vectors $\mathbf e_m$ and $\mathbf e_{m'}$, and the sum $i$ runs over the lattice-positions $\mathbf R_i$ of the individual spins; the notation is analogous to reference \cite{ruckriegelandkopietz2015}.

Knowing the phonon-bath spectral density \eqref{eq:J-ph}, we then find a low-frequency approximation $J_{\mathrm{lf}, mm'}(\epsilon)$ and, in turn, determine the phonon-bath-induced spin inertia from the difference $J_{mm'}(\epsilon) - J_{\mathrm{lf},mm'}(\epsilon)$ as in Eq. \eqref{eq:inertia}.

To be specific, let us consider a layer of yttrium iron garnet (YIG) that is sandwiched between two bulk layers of gadolinium gallium garnet (GGG); afterwards, we consider implications for a layer of YIG on top of a bulk of GGG. We model the system as a planar spin lattice (in the $z=0$ plane) that is embedded in a three-dimensional lattice with lattice constant $a$. In this geometry, only phonons travelling in $z$-direction contribute to the macrospin damping and inertia. Focusing on acoustic phonons, we use the dispersion relation for a simple cubic lattice $\omega_{k_z \lambda} = (2/a)\, v_\lambda \left|\sin (k_z a/2)\right|$, where $v_\lambda$ is the sound velocity for transversal ($\lambda \in \lbrace x,y \rbrace$) or longitudinal ($\lambda=z$) polarization. At low energies $\epsilon$ the bath spectral density \eqref{eq:J-ph} is governed by long wavelength (small $\mathbf k$) acoustic phonons, which have an approximately linear dispersion relation, $\omega_{k_z \lambda} = v_\lambda \left|k_z\right|$. In turn, $J_{mm'}(\epsilon)$ becomes linear in $\epsilon$ at low energies and we can use the low-frequency approximation $J_{\mathrm{lf}, mm'}(\epsilon) = \alpha_{mm'}\, \epsilon\, \Theta(\epsilon)$ with a tensorial Gilbert-damping coefficient $\alpha_{mm'}$; for YIG on GGG, a tensorial Gilbert damping has been found before \cite{streib2018damping}. The phonon-bath-induced spin inertia is now found from the high-frequency modes as in Eq. \eqref{eq:inertia}; it also takes a tensorial form $I_{mm'} = (2/\pi) \int_{-\infty}^{\infty}\! d\epsilon\, [J_{mm'}(\epsilon) - J_{\mathrm{lf},mm'}(\epsilon)]/\epsilon^3$. Explicitly, we find the Gilbert-damping coefficient $\alpha_{mm'} = \delta_{mm'} (1+\delta_{mz})^2 B_{mz}^2 a/2 M S_1 S v_m^3$ and the phonon-bath-induced spin inertia
\begin{equation}
I_{mm'} = \frac{0.9}{\pi}\, \frac{a}{v_m}\, \alpha_{mm'}\ , \label{eq:ph-inertia}
\end{equation}
where, after making the $\epsilon$-integral dimensionless, we evaluated and approximated it numerically to $0.9$; a detailed calculation is provided in the Supplemental Material. For YIG on top of GGG (not sandwiched) only about half the phonon modes are present, such that we expect the Gilbert-damping coefficient, and with it the spin inertia, to be only half as large.

Bath-induced spin inertia and Gilbert damping are closely related, as becomes clear from Eq. \eqref{eq:ph-inertia}; namely, the spin inertia is proportional to the Gilbert-damping coefficient. Thus, since the Gilbert-damping tensor $\alpha_{mm'}$ is diagonal, also the spin-inertia tensor $I_{mm'}$ is diagonal. The proportionality constant $\tau_m = I_{mm}/\alpha_{mm} = 0.9\, a/\pi v_m$ is typically used in experimental works to characterize spin inertia or nutation dynamics \cite{li2015, neeraj2021, unikandanunni2022}. For the GGG phonon bath, with lattice spacing $a = 1.2383\, \mathrm{nm}$ \cite{FUJII20013666} and phonon velocities $v_z = 6411\, \mathrm{m}/\mathrm{s}$ (longitudinal) and $v_x= v_y = 3568\, \mathrm{m}/\mathrm{s}$ (transversal) \cite{kleszczewski1988phonon, streib2018damping}, we find $\tau_z \approx 55\, \mathrm{fs}$ and $\tau_x = \tau_y \approx 99.4\, \mathrm{fs}$. These results, compared to previous measurements on metallic magnets, are roughly in the same order of magnitude \cite{li2015, unikandanunni2022} or about two to three orders of magnitude lower \cite{neeraj2021}. 

Note that the frequency of the nutation resonance peak may deviate from previous experimental results on metallic magnets, as it also depends on the Gilbert-damping coefficient \cite{olive2015deviation}. Using the effective (oscillating) mass of GGG $M= \rho\, a^3$ with the GGG density $\rho = 7.07\cdot 10^3\, \mathrm{kg}/\mathrm{m}^3$ \cite{kleszczewski1988phonon, streib2018damping}, the magnetoelastic coupling coefficients $B_{zz} = 6.6\cdot 10^{-22}\, \mathrm{J}$ and $B_{xz}= B_{yz} = 13.2\cdot 10^{-22}\, \mathrm{J}$ with the YIG-lattice-site spin length $S_1 = 14.2\, \hbar$ \cite{ruckriegel2014magnetoelastic}, we find $\alpha_{zz} = 2\cdot 10^{-7}/S$ and $\alpha_{xx} = \alpha_{yy} = 1.2 \cdot 10^{-6}/S$ for the elements of the tensorial Gilbert-damping coefficient. For those numbers, using the ``weak coupling'' approximation of reference \cite{olive2015deviation}, the nutation resonance peaks would be in the order of X-ray frequencies. While probably impractical in for experiments, it does not affect the illustration purposes of our simple example.

\textit{Discussion and Conclusion.}—Using the Caldeira-Leggett approach, we have shown that the high-frequency modes of an environment (bath) should—universally—lead to bath-induced spin inertia. The low-frequency bath modes, if Ohmic, lead to the usual Gilbert-damping term. This has two important consequences. First, our results suggest that the appearance of bath-induced spin inertia is more robust (or universal) than the form of Gilbert damping; while for non-Ohmic baths Gilbert damping turns into fractional Gilbert damping \cite{verstraten2022}, the bath-induced spin inertia retains its form. Second, our results suggest that, whenever a dissipation channel (environment/bath) is added to some spin system, the spin dynamics will also acquire an additional contribution to its spin inertia. In short, spin relaxation is always accompanied by spin inertia.

To show how our general derivation based on the Caldeira-Leggett approach applies to more realistic models, we considered the dynamics of a macrospin in a phonon bath; specifically, we considered a heterostructure based on YIG and GGG. We expect that our approach can be applied, along similar lines, to many other spin environments as well; for example, to electron baths in metallic magnets and to metallic leads in contact with insulating magnets. Furthermore, we believe that our approach can be generalized from the macrospin case considered here to the more general case of spin lattices and, in turn, also to continuum theories of spin textures.

Because the high-frequency bath modes affect the action, Eq. \eqref{eq:action}, already before the quasi-classical approximation, we also expect small-spin systems, as investigated in \cite{zhukov2018spin, smirnov2018theory}, to be affected by bath-induced spin inertia from high-frequency bath modes.

\begin{acknowledgments}
\textit{Acknowledgements.}—We thank M.~Cherkasskii, E.~Di~Salvo, A.~Kamra, A.~Semisalova, A.~Shnirman, D.~Thonig, R.~Willa, and X.~R.~Wang for fruitful discussions. This work is part of the research programme Fluid Spintronics with project number 182.069, financed by the Dutch Research Council (NWO). R.~C.~V. is supported by (NWO, Grant No. 680.92.18.05). R.~A.~D. has received funding from the European Research Council (ERC) under the European Union’s Horizon 2020 research and innovation programme (Grant No. 725509).
\end{acknowledgments}

\bibliography{reference.bib}

\begin{thebibliography}{52}%
\makeatletter
\providecommand \@ifxundefined [1]{%
 \@ifx{#1\undefined}
}%
\providecommand \@ifnum [1]{%
 \ifnum #1\expandafter \@firstoftwo
 \else \expandafter \@secondoftwo
 \fi
}%
\providecommand \@ifx [1]{%
 \ifx #1\expandafter \@firstoftwo
 \else \expandafter \@secondoftwo
 \fi
}%
\providecommand \natexlab [1]{#1}%
\providecommand \enquote  [1]{``#1''}%
\providecommand \bibnamefont  [1]{#1}%
\providecommand \bibfnamefont [1]{#1}%
\providecommand \citenamefont [1]{#1}%
\providecommand \href@noop [0]{\@secondoftwo}%
\providecommand \href [0]{\begingroup \@sanitize@url \@href}%
\providecommand \@href[1]{\@@startlink{#1}\@@href}%
\providecommand \@@href[1]{\endgroup#1\@@endlink}%
\providecommand \@sanitize@url [0]{\catcode `\\12\catcode `\$12\catcode
  `\&12\catcode `\#12\catcode `\^12\catcode `\_12\catcode `\%12\relax}%
\providecommand \@@startlink[1]{}%
\providecommand \@@endlink[0]{}%
\providecommand \url  [0]{\begingroup\@sanitize@url \@url }%
\providecommand \@url [1]{\endgroup\@href {#1}{\urlprefix }}%
\providecommand \urlprefix  [0]{URL }%
\providecommand \Eprint [0]{\href }%
\providecommand \doibase [0]{https://doi.org/}%
\providecommand \selectlanguage [0]{\@gobble}%
\providecommand \bibinfo  [0]{\@secondoftwo}%
\providecommand \bibfield  [0]{\@secondoftwo}%
\providecommand \translation [1]{[#1]}%
\providecommand \BibitemOpen [0]{}%
\providecommand \bibitemStop [0]{}%
\providecommand \bibitemNoStop [0]{.\EOS\space}%
\providecommand \EOS [0]{\spacefactor3000\relax}%
\providecommand \BibitemShut  [1]{\csname bibitem#1\endcsname}%
\let\auto@bib@innerbib\@empty
%</preamble>
\bibitem [{\citenamefont {Bloch}(1946)}]{bloch1946}%
  \BibitemOpen
  \bibfield  {author} {\bibinfo {author} {\bibfnamefont {F.}~\bibnamefont
  {Bloch}},\ }\bibfield  {title} {\bibinfo {title} {Nuclear induction},\
  }\href@noop {} {\bibfield  {journal} {\bibinfo  {journal} {Physical review}\
  }\textbf {\bibinfo {volume} {70}},\ \bibinfo {pages} {460} (\bibinfo {year}
  {1946})}\BibitemShut {NoStop}%
\bibitem [{\citenamefont {Lifshitz}\ and\ \citenamefont
  {Landau}(1935)}]{lifshitz1935}%
  \BibitemOpen
  \bibfield  {author} {\bibinfo {author} {\bibfnamefont {E.}~\bibnamefont
  {Lifshitz}}\ and\ \bibinfo {author} {\bibfnamefont {L.}~\bibnamefont
  {Landau}},\ }\bibfield  {title} {\bibinfo {title} {On the theory of the
  dispersion of magnetic permeability in ferromagnetic bodies},\ }\href@noop {}
  {\bibfield  {journal} {\bibinfo  {journal} {Phys. Z. Sowjetunion}\ }\textbf
  {\bibinfo {volume} {8}} (\bibinfo {year} {1935})}\BibitemShut {NoStop}%
\bibitem [{\citenamefont {Landau}\ and\ \citenamefont
  {Lifshitz}(1992)}]{landau1992}%
  \BibitemOpen
  \bibfield  {author} {\bibinfo {author} {\bibfnamefont {L.}~\bibnamefont
  {Landau}}\ and\ \bibinfo {author} {\bibfnamefont {E.}~\bibnamefont
  {Lifshitz}},\ }\bibfield  {title} {\bibinfo {title} {On the theory of the
  dispersion of magnetic permeability in ferromagnetic bodies},\ }in\
  \href@noop {} {\emph {\bibinfo {booktitle} {Perspectives in Theoretical
  Physics}}}\ (\bibinfo  {publisher} {Elsevier},\ \bibinfo {year} {1992})\ pp.\
  \bibinfo {pages} {51--65}\BibitemShut {NoStop}%
\bibitem [{\citenamefont {Gilbert}(1956)}]{gilbert1956}%
  \BibitemOpen
  \bibfield  {author} {\bibinfo {author} {\bibfnamefont {T.~L.}\ \bibnamefont
  {Gilbert}},\ }\bibfield  {title} {\bibinfo {title} {Formulation, foundations
  and applications of the phenomenological theory of ferromagnetism.},\
  }\href@noop {} {\bibfield  {journal} {\bibinfo  {journal} {Ph. D. Thesis}\ }
  (\bibinfo {year} {1956})}\BibitemShut {NoStop}%
\bibitem [{\citenamefont {Gilbert}(2004)}]{gilbert2004}%
  \BibitemOpen
  \bibfield  {author} {\bibinfo {author} {\bibfnamefont {T.~L.}\ \bibnamefont
  {Gilbert}},\ }\bibfield  {title} {\bibinfo {title} {A phenomenological theory
  of damping in ferromagnetic materials},\ }\href@noop {} {\bibfield  {journal}
  {\bibinfo  {journal} {IEEE transactions on magnetics}\ }\textbf {\bibinfo
  {volume} {40}},\ \bibinfo {pages} {3443} (\bibinfo {year}
  {2004})}\BibitemShut {NoStop}%
\bibitem [{\citenamefont {Neeraj}\ \emph {et~al.}(2021)\citenamefont {Neeraj},
  \citenamefont {Awari}, \citenamefont {Kovalev}, \citenamefont {Polley},
  \citenamefont {Zhou~Hagstr{\"o}m}, \citenamefont {Arekapudi}, \citenamefont
  {Semisalova}, \citenamefont {Lenz}, \citenamefont {Green}, \citenamefont
  {Deinert} \emph {et~al.}}]{neeraj2021}%
  \BibitemOpen
  \bibfield  {author} {\bibinfo {author} {\bibfnamefont {K.}~\bibnamefont
  {Neeraj}}, \bibinfo {author} {\bibfnamefont {N.}~\bibnamefont {Awari}},
  \bibinfo {author} {\bibfnamefont {S.}~\bibnamefont {Kovalev}}, \bibinfo
  {author} {\bibfnamefont {D.}~\bibnamefont {Polley}}, \bibinfo {author}
  {\bibfnamefont {N.}~\bibnamefont {Zhou~Hagstr{\"o}m}}, \bibinfo {author}
  {\bibfnamefont {S.~S. P.~K.}\ \bibnamefont {Arekapudi}}, \bibinfo {author}
  {\bibfnamefont {A.}~\bibnamefont {Semisalova}}, \bibinfo {author}
  {\bibfnamefont {K.}~\bibnamefont {Lenz}}, \bibinfo {author} {\bibfnamefont
  {B.}~\bibnamefont {Green}}, \bibinfo {author} {\bibfnamefont {J.-C.}\
  \bibnamefont {Deinert}}, \emph {et~al.},\ }\bibfield  {title} {\bibinfo
  {title} {Inertial spin dynamics in ferromagnets},\ }\href@noop {} {\bibfield
  {journal} {\bibinfo  {journal} {Nature Physics}\ }\textbf {\bibinfo {volume}
  {17}},\ \bibinfo {pages} {245} (\bibinfo {year} {2021})}\BibitemShut
  {NoStop}%
\bibitem [{\citenamefont {Unikandanunni}\ \emph {et~al.}(2022)\citenamefont
  {Unikandanunni}, \citenamefont {Medapalli}, \citenamefont {Asa},
  \citenamefont {Albisetti}, \citenamefont {Petti}, \citenamefont {Bertacco},
  \citenamefont {Fullerton},\ and\ \citenamefont
  {Bonetti}}]{unikandanunni2022}%
  \BibitemOpen
  \bibfield  {author} {\bibinfo {author} {\bibfnamefont {V.}~\bibnamefont
  {Unikandanunni}}, \bibinfo {author} {\bibfnamefont {R.}~\bibnamefont
  {Medapalli}}, \bibinfo {author} {\bibfnamefont {M.}~\bibnamefont {Asa}},
  \bibinfo {author} {\bibfnamefont {E.}~\bibnamefont {Albisetti}}, \bibinfo
  {author} {\bibfnamefont {D.}~\bibnamefont {Petti}}, \bibinfo {author}
  {\bibfnamefont {R.}~\bibnamefont {Bertacco}}, \bibinfo {author}
  {\bibfnamefont {E.~E.}\ \bibnamefont {Fullerton}},\ and\ \bibinfo {author}
  {\bibfnamefont {S.}~\bibnamefont {Bonetti}},\ }\bibfield  {title} {\bibinfo
  {title} {Inertial spin dynamics in epitaxial cobalt films},\ }\href@noop {}
  {\bibfield  {journal} {\bibinfo  {journal} {Physical review letters}\
  }\textbf {\bibinfo {volume} {129}},\ \bibinfo {pages} {237201} (\bibinfo
  {year} {2022})}\BibitemShut {NoStop}%
\bibitem [{\citenamefont {Li}\ \emph {et~al.}(2015)\citenamefont {Li},
  \citenamefont {Barra}, \citenamefont {Auffret}, \citenamefont {Ebels},\ and\
  \citenamefont {Bailey}}]{li2015}%
  \BibitemOpen
  \bibfield  {author} {\bibinfo {author} {\bibfnamefont {Y.}~\bibnamefont
  {Li}}, \bibinfo {author} {\bibfnamefont {A.-L.}\ \bibnamefont {Barra}},
  \bibinfo {author} {\bibfnamefont {S.}~\bibnamefont {Auffret}}, \bibinfo
  {author} {\bibfnamefont {U.}~\bibnamefont {Ebels}},\ and\ \bibinfo {author}
  {\bibfnamefont {W.~E.}\ \bibnamefont {Bailey}},\ }\bibfield  {title}
  {\bibinfo {title} {Inertial terms to magnetization dynamics in ferromagnetic
  thin films},\ }\href@noop {} {\bibfield  {journal} {\bibinfo  {journal}
  {Physical Review B}\ }\textbf {\bibinfo {volume} {92}},\ \bibinfo {pages}
  {140413} (\bibinfo {year} {2015})}\BibitemShut {NoStop}%
\bibitem [{\citenamefont {Cherkasskii}\ \emph {et~al.}(2022)\citenamefont
  {Cherkasskii}, \citenamefont {Barsukov}, \citenamefont {Mondal},
  \citenamefont {Farle},\ and\ \citenamefont
  {Semisalova}}]{cherkasskii2022theory}%
  \BibitemOpen
  \bibfield  {author} {\bibinfo {author} {\bibfnamefont {M.}~\bibnamefont
  {Cherkasskii}}, \bibinfo {author} {\bibfnamefont {I.}~\bibnamefont
  {Barsukov}}, \bibinfo {author} {\bibfnamefont {R.}~\bibnamefont {Mondal}},
  \bibinfo {author} {\bibfnamefont {M.}~\bibnamefont {Farle}},\ and\ \bibinfo
  {author} {\bibfnamefont {A.}~\bibnamefont {Semisalova}},\ }\bibfield  {title}
  {\bibinfo {title} {Theory of inertial spin dynamics in anisotropic
  ferromagnets},\ }\href@noop {} {\bibfield  {journal} {\bibinfo  {journal}
  {Physical Review B}\ }\textbf {\bibinfo {volume} {106}},\ \bibinfo {pages}
  {054428} (\bibinfo {year} {2022})}\BibitemShut {NoStop}%
\bibitem [{\citenamefont {Cherkasskii}\ \emph {et~al.}(2020)\citenamefont
  {Cherkasskii}, \citenamefont {Farle},\ and\ \citenamefont
  {Semisalova}}]{cherkasskii2020nutation}%
  \BibitemOpen
  \bibfield  {author} {\bibinfo {author} {\bibfnamefont {M.}~\bibnamefont
  {Cherkasskii}}, \bibinfo {author} {\bibfnamefont {M.}~\bibnamefont {Farle}},\
  and\ \bibinfo {author} {\bibfnamefont {A.}~\bibnamefont {Semisalova}},\
  }\bibfield  {title} {\bibinfo {title} {Nutation resonance in ferromagnets},\
  }\href@noop {} {\bibfield  {journal} {\bibinfo  {journal} {Physical Review
  B}\ }\textbf {\bibinfo {volume} {102}},\ \bibinfo {pages} {184432} (\bibinfo
  {year} {2020})}\BibitemShut {NoStop}%
\bibitem [{\citenamefont {Mondal}\ and\ \citenamefont
  {R{\'o}zsa}(2022)}]{mondal2022inertial}%
  \BibitemOpen
  \bibfield  {author} {\bibinfo {author} {\bibfnamefont {R.}~\bibnamefont
  {Mondal}}\ and\ \bibinfo {author} {\bibfnamefont {L.}~\bibnamefont
  {R{\'o}zsa}},\ }\bibfield  {title} {\bibinfo {title} {Inertial spin waves in
  ferromagnets and antiferromagnets},\ }\href@noop {} {\bibfield  {journal}
  {\bibinfo  {journal} {Physical Review B}\ }\textbf {\bibinfo {volume}
  {106}},\ \bibinfo {pages} {134422} (\bibinfo {year} {2022})}\BibitemShut
  {NoStop}%
\bibitem [{\citenamefont {Cherkasskii}\ \emph {et~al.}(2021)\citenamefont
  {Cherkasskii}, \citenamefont {Farle},\ and\ \citenamefont
  {Semisalova}}]{cherkasskii2021dispersion}%
  \BibitemOpen
  \bibfield  {author} {\bibinfo {author} {\bibfnamefont {M.}~\bibnamefont
  {Cherkasskii}}, \bibinfo {author} {\bibfnamefont {M.}~\bibnamefont {Farle}},\
  and\ \bibinfo {author} {\bibfnamefont {A.}~\bibnamefont {Semisalova}},\
  }\bibfield  {title} {\bibinfo {title} {Dispersion relation of nutation
  surface spin waves in ferromagnets},\ }\href@noop {} {\bibfield  {journal}
  {\bibinfo  {journal} {Physical Review B}\ }\textbf {\bibinfo {volume}
  {103}},\ \bibinfo {pages} {174435} (\bibinfo {year} {2021})}\BibitemShut
  {NoStop}%
\bibitem [{\citenamefont {Titov}\ \emph {et~al.}(2022)\citenamefont {Titov},
  \citenamefont {Dowling}, \citenamefont {Kalmykov},\ and\ \citenamefont
  {Cherkasskii}}]{titov2022nutation}%
  \BibitemOpen
  \bibfield  {author} {\bibinfo {author} {\bibfnamefont {S.~V.}\ \bibnamefont
  {Titov}}, \bibinfo {author} {\bibfnamefont {W.~J.}\ \bibnamefont {Dowling}},
  \bibinfo {author} {\bibfnamefont {Y.~P.}\ \bibnamefont {Kalmykov}},\ and\
  \bibinfo {author} {\bibfnamefont {M.}~\bibnamefont {Cherkasskii}},\
  }\bibfield  {title} {\bibinfo {title} {Nutation spin waves in ferromagnets},\
  }\href@noop {} {\bibfield  {journal} {\bibinfo  {journal} {Physical Review
  B}\ }\textbf {\bibinfo {volume} {105}},\ \bibinfo {pages} {214414} (\bibinfo
  {year} {2022})}\BibitemShut {NoStop}%
\bibitem [{\citenamefont {F{\"a}hnle}(2019)}]{fahnle2019}%
  \BibitemOpen
  \bibfield  {author} {\bibinfo {author} {\bibfnamefont {M.}~\bibnamefont
  {F{\"a}hnle}},\ }\bibfield  {title} {\bibinfo {title} {Comparison of theories
  of fast and ultrafast magnetization dynamics},\ }\href@noop {} {\bibfield
  {journal} {\bibinfo  {journal} {Journal of Magnetism and Magnetic Materials}\
  }\textbf {\bibinfo {volume} {469}},\ \bibinfo {pages} {28} (\bibinfo {year}
  {2019})}\BibitemShut {NoStop}%
\bibitem [{\citenamefont {Kimel}\ \emph {et~al.}(2009)\citenamefont {Kimel},
  \citenamefont {Ivanov}, \citenamefont {Pisarev}, \citenamefont {Usachev},
  \citenamefont {Kirilyuk},\ and\ \citenamefont {Rasing}}]{kimel2009}%
  \BibitemOpen
  \bibfield  {author} {\bibinfo {author} {\bibfnamefont {A.}~\bibnamefont
  {Kimel}}, \bibinfo {author} {\bibfnamefont {B.}~\bibnamefont {Ivanov}},
  \bibinfo {author} {\bibfnamefont {R.}~\bibnamefont {Pisarev}}, \bibinfo
  {author} {\bibfnamefont {P.}~\bibnamefont {Usachev}}, \bibinfo {author}
  {\bibfnamefont {A.}~\bibnamefont {Kirilyuk}},\ and\ \bibinfo {author}
  {\bibfnamefont {T.}~\bibnamefont {Rasing}},\ }\bibfield  {title} {\bibinfo
  {title} {Inertia-driven spin switching in antiferromagnets},\ }\href@noop {}
  {\bibfield  {journal} {\bibinfo  {journal} {Nature Physics}\ }\textbf
  {\bibinfo {volume} {5}},\ \bibinfo {pages} {727} (\bibinfo {year}
  {2009})}\BibitemShut {NoStop}%
\bibitem [{\citenamefont {Mondal}\ \emph {et~al.}(2023)\citenamefont {Mondal},
  \citenamefont {R{\'o}zsa}, \citenamefont {Farle}, \citenamefont {Oppeneer},
  \citenamefont {Nowak},\ and\ \citenamefont
  {Cherkasskii}}]{mondal2023inertial}%
  \BibitemOpen
  \bibfield  {author} {\bibinfo {author} {\bibfnamefont {R.}~\bibnamefont
  {Mondal}}, \bibinfo {author} {\bibfnamefont {L.}~\bibnamefont {R{\'o}zsa}},
  \bibinfo {author} {\bibfnamefont {M.}~\bibnamefont {Farle}}, \bibinfo
  {author} {\bibfnamefont {P.~M.}\ \bibnamefont {Oppeneer}}, \bibinfo {author}
  {\bibfnamefont {U.}~\bibnamefont {Nowak}},\ and\ \bibinfo {author}
  {\bibfnamefont {M.}~\bibnamefont {Cherkasskii}},\ }\bibfield  {title}
  {\bibinfo {title} {Inertial effects in ultrafast spin dynamics},\ }\href@noop
  {} {\bibfield  {journal} {\bibinfo  {journal} {Journal of Magnetism and
  Magnetic Materials}\ ,\ \bibinfo {pages} {170830}} (\bibinfo {year}
  {2023})}\BibitemShut {NoStop}%
\bibitem [{\citenamefont {Suhl}(1998)}]{suhl1998theory}%
  \BibitemOpen
  \bibfield  {author} {\bibinfo {author} {\bibfnamefont {H.}~\bibnamefont
  {Suhl}},\ }\bibfield  {title} {\bibinfo {title} {Theory of the magnetic
  damping constant},\ }\href@noop {} {\bibfield  {journal} {\bibinfo  {journal}
  {IEEE Transactions on magnetics}\ }\textbf {\bibinfo {volume} {34}},\
  \bibinfo {pages} {1834} (\bibinfo {year} {1998})}\BibitemShut {NoStop}%
\bibitem [{\citenamefont {Zhu}\ \emph {et~al.}(2004)\citenamefont {Zhu},
  \citenamefont {Nussinov}, \citenamefont {Shnirman},\ and\ \citenamefont
  {Balatsky}}]{zhu2004novel}%
  \BibitemOpen
  \bibfield  {author} {\bibinfo {author} {\bibfnamefont {J.-X.}\ \bibnamefont
  {Zhu}}, \bibinfo {author} {\bibfnamefont {Z.}~\bibnamefont {Nussinov}},
  \bibinfo {author} {\bibfnamefont {A.}~\bibnamefont {Shnirman}},\ and\
  \bibinfo {author} {\bibfnamefont {A.~V.}\ \bibnamefont {Balatsky}},\
  }\bibfield  {title} {\bibinfo {title} {Novel spin dynamics in a josephson
  junction},\ }\href@noop {} {\bibfield  {journal} {\bibinfo  {journal}
  {Physical review letters}\ }\textbf {\bibinfo {volume} {92}},\ \bibinfo
  {pages} {107001} (\bibinfo {year} {2004})}\BibitemShut {NoStop}%
\bibitem [{\citenamefont {Ciornei}\ \emph {et~al.}(2011)\citenamefont
  {Ciornei}, \citenamefont {Rub{\'\i}},\ and\ \citenamefont
  {Wegrowe}}]{ciornei2011}%
  \BibitemOpen
  \bibfield  {author} {\bibinfo {author} {\bibfnamefont {M.-C.}\ \bibnamefont
  {Ciornei}}, \bibinfo {author} {\bibfnamefont {J.}~\bibnamefont {Rub{\'\i}}},\
  and\ \bibinfo {author} {\bibfnamefont {J.-E.}\ \bibnamefont {Wegrowe}},\
  }\bibfield  {title} {\bibinfo {title} {Magnetization dynamics in the inertial
  regime: Nutation predicted at short time scales},\ }\href@noop {} {\bibfield
  {journal} {\bibinfo  {journal} {Physical Review B}\ }\textbf {\bibinfo
  {volume} {83}},\ \bibinfo {pages} {020410} (\bibinfo {year}
  {2011})}\BibitemShut {NoStop}%
\bibitem [{\citenamefont {Wegrowe}\ and\ \citenamefont
  {Ciornei}(2012)}]{wegrowe2012}%
  \BibitemOpen
  \bibfield  {author} {\bibinfo {author} {\bibfnamefont {J.-E.}\ \bibnamefont
  {Wegrowe}}\ and\ \bibinfo {author} {\bibfnamefont {M.-C.}\ \bibnamefont
  {Ciornei}},\ }\bibfield  {title} {\bibinfo {title} {Magnetization dynamics,
  gyromagnetic relation, and inertial effects},\ }\href@noop {} {\bibfield
  {journal} {\bibinfo  {journal} {American Journal of Physics}\ }\textbf
  {\bibinfo {volume} {80}},\ \bibinfo {pages} {607} (\bibinfo {year}
  {2012})}\BibitemShut {NoStop}%
\bibitem [{\citenamefont {F{\"a}hnle}\ \emph {et~al.}(2011)\citenamefont
  {F{\"a}hnle}, \citenamefont {Steiauf},\ and\ \citenamefont
  {Illg}}]{fahnle2011}%
  \BibitemOpen
  \bibfield  {author} {\bibinfo {author} {\bibfnamefont {M.}~\bibnamefont
  {F{\"a}hnle}}, \bibinfo {author} {\bibfnamefont {D.}~\bibnamefont
  {Steiauf}},\ and\ \bibinfo {author} {\bibfnamefont {C.}~\bibnamefont
  {Illg}},\ }\bibfield  {title} {\bibinfo {title} {Generalized gilbert equation
  including inertial damping: Derivation from an extended breathing fermi
  surface model},\ }\href@noop {} {\bibfield  {journal} {\bibinfo  {journal}
  {Physical Review B}\ }\textbf {\bibinfo {volume} {84}},\ \bibinfo {pages}
  {172403} (\bibinfo {year} {2011})}\BibitemShut {NoStop}%
\bibitem [{\citenamefont {Bhattacharjee}\ \emph {et~al.}(2012)\citenamefont
  {Bhattacharjee}, \citenamefont {Nordstr{\"o}m},\ and\ \citenamefont
  {Fransson}}]{bhattacharjee2012}%
  \BibitemOpen
  \bibfield  {author} {\bibinfo {author} {\bibfnamefont {S.}~\bibnamefont
  {Bhattacharjee}}, \bibinfo {author} {\bibfnamefont {L.}~\bibnamefont
  {Nordstr{\"o}m}},\ and\ \bibinfo {author} {\bibfnamefont {J.}~\bibnamefont
  {Fransson}},\ }\bibfield  {title} {\bibinfo {title} {Atomistic spin dynamic
  method with both damping and moment of inertia effects included from first
  principles},\ }\href@noop {} {\bibfield  {journal} {\bibinfo  {journal}
  {Physical review letters}\ }\textbf {\bibinfo {volume} {108}},\ \bibinfo
  {pages} {057204} (\bibinfo {year} {2012})}\BibitemShut {NoStop}%
\bibitem [{\citenamefont {Kikuchi}\ and\ \citenamefont
  {Tatara}(2015)}]{kikuchi2015}%
  \BibitemOpen
  \bibfield  {author} {\bibinfo {author} {\bibfnamefont {T.}~\bibnamefont
  {Kikuchi}}\ and\ \bibinfo {author} {\bibfnamefont {G.}~\bibnamefont
  {Tatara}},\ }\bibfield  {title} {\bibinfo {title} {Spin dynamics with inertia
  in metallic ferromagnets},\ }\href@noop {} {\bibfield  {journal} {\bibinfo
  {journal} {Physical Review B}\ }\textbf {\bibinfo {volume} {92}},\ \bibinfo
  {pages} {184410} (\bibinfo {year} {2015})}\BibitemShut {NoStop}%
\bibitem [{\citenamefont {Hurst}\ \emph {et~al.}(2020)\citenamefont {Hurst},
  \citenamefont {Galitski},\ and\ \citenamefont {Heikkil{\"a}}}]{hurst2020}%
  \BibitemOpen
  \bibfield  {author} {\bibinfo {author} {\bibfnamefont {H.~M.}\ \bibnamefont
  {Hurst}}, \bibinfo {author} {\bibfnamefont {V.}~\bibnamefont {Galitski}},\
  and\ \bibinfo {author} {\bibfnamefont {T.~T.}\ \bibnamefont {Heikkil{\"a}}},\
  }\bibfield  {title} {\bibinfo {title} {Electron-induced massive dynamics of
  magnetic domain walls},\ }\href@noop {} {\bibfield  {journal} {\bibinfo
  {journal} {Physical Review B}\ }\textbf {\bibinfo {volume} {101}},\ \bibinfo
  {pages} {054407} (\bibinfo {year} {2020})}\BibitemShut {NoStop}%
\bibitem [{\citenamefont {Mondal}\ \emph {et~al.}(2017)\citenamefont {Mondal},
  \citenamefont {Berritta}, \citenamefont {Nandy},\ and\ \citenamefont
  {Oppeneer}}]{mondal2017}%
  \BibitemOpen
  \bibfield  {author} {\bibinfo {author} {\bibfnamefont {R.}~\bibnamefont
  {Mondal}}, \bibinfo {author} {\bibfnamefont {M.}~\bibnamefont {Berritta}},
  \bibinfo {author} {\bibfnamefont {A.~K.}\ \bibnamefont {Nandy}},\ and\
  \bibinfo {author} {\bibfnamefont {P.~M.}\ \bibnamefont {Oppeneer}},\
  }\bibfield  {title} {\bibinfo {title} {Relativistic theory of magnetic
  inertia in ultrafast spin dynamics},\ }\href@noop {} {\bibfield  {journal}
  {\bibinfo  {journal} {Physical Review B}\ }\textbf {\bibinfo {volume} {96}},\
  \bibinfo {pages} {024425} (\bibinfo {year} {2017})}\BibitemShut {NoStop}%
\bibitem [{\citenamefont {Mondal}\ \emph {et~al.}(2018)\citenamefont {Mondal},
  \citenamefont {Berritta},\ and\ \citenamefont {Oppeneer}}]{mondal2018}%
  \BibitemOpen
  \bibfield  {author} {\bibinfo {author} {\bibfnamefont {R.}~\bibnamefont
  {Mondal}}, \bibinfo {author} {\bibfnamefont {M.}~\bibnamefont {Berritta}},\
  and\ \bibinfo {author} {\bibfnamefont {P.~M.}\ \bibnamefont {Oppeneer}},\
  }\bibfield  {title} {\bibinfo {title} {Generalisation of gilbert damping and
  magnetic inertia parameter as a series of higher-order relativistic terms},\
  }\href@noop {} {\bibfield  {journal} {\bibinfo  {journal} {Journal of
  Physics: Condensed Matter}\ }\textbf {\bibinfo {volume} {30}},\ \bibinfo
  {pages} {265801} (\bibinfo {year} {2018})}\BibitemShut {NoStop}%
\bibitem [{\citenamefont {Anders}\ \emph {et~al.}(2022)\citenamefont {Anders},
  \citenamefont {Sait},\ and\ \citenamefont {Horsley}}]{anders2022quantum}%
  \BibitemOpen
  \bibfield  {author} {\bibinfo {author} {\bibfnamefont {J.}~\bibnamefont
  {Anders}}, \bibinfo {author} {\bibfnamefont {C.~R.}\ \bibnamefont {Sait}},\
  and\ \bibinfo {author} {\bibfnamefont {S.~A.}\ \bibnamefont {Horsley}},\
  }\bibfield  {title} {\bibinfo {title} {Quantum brownian motion for magnets},\
  }\href@noop {} {\bibfield  {journal} {\bibinfo  {journal} {New Journal of
  Physics}\ }\textbf {\bibinfo {volume} {24}},\ \bibinfo {pages} {033020}
  (\bibinfo {year} {2022})}\BibitemShut {NoStop}%
\bibitem [{\citenamefont {Titov}\ \emph {et~al.}(2021)\citenamefont {Titov},
  \citenamefont {Coffey}, \citenamefont {Kalmykov}, \citenamefont {Zarifakis},\
  and\ \citenamefont {Titov}}]{titov2021}%
  \BibitemOpen
  \bibfield  {author} {\bibinfo {author} {\bibfnamefont {S.}~\bibnamefont
  {Titov}}, \bibinfo {author} {\bibfnamefont {W.}~\bibnamefont {Coffey}},
  \bibinfo {author} {\bibfnamefont {Y.~P.}\ \bibnamefont {Kalmykov}}, \bibinfo
  {author} {\bibfnamefont {M.}~\bibnamefont {Zarifakis}},\ and\ \bibinfo
  {author} {\bibfnamefont {A.}~\bibnamefont {Titov}},\ }\bibfield  {title}
  {\bibinfo {title} {Inertial magnetization dynamics of ferromagnetic
  nanoparticles including thermal agitation},\ }\href@noop {} {\bibfield
  {journal} {\bibinfo  {journal} {Physical Review B}\ }\textbf {\bibinfo
  {volume} {103}},\ \bibinfo {pages} {144433} (\bibinfo {year}
  {2021})}\BibitemShut {NoStop}%
\bibitem [{\citenamefont {Caldeira}\ and\ \citenamefont
  {Leggett}(1983{\natexlab{a}})}]{caldeiraandleggett1983}%
  \BibitemOpen
  \bibfield  {author} {\bibinfo {author} {\bibfnamefont {A.~O.}\ \bibnamefont
  {Caldeira}}\ and\ \bibinfo {author} {\bibfnamefont {A.~J.}\ \bibnamefont
  {Leggett}},\ }\bibfield  {title} {\bibinfo {title} {Quantum tunnelling in a
  dissipative system},\ }\href@noop {} {\bibfield  {journal} {\bibinfo
  {journal} {Annals of physics}\ }\textbf {\bibinfo {volume} {149}},\ \bibinfo
  {pages} {374} (\bibinfo {year} {1983}{\natexlab{a}})}\BibitemShut {NoStop}%
\bibitem [{\citenamefont {Caldeira}\ and\ \citenamefont
  {Leggett}(1983{\natexlab{b}})}]{caldeiraandleggett1983a}%
  \BibitemOpen
  \bibfield  {author} {\bibinfo {author} {\bibfnamefont {A.~O.}\ \bibnamefont
  {Caldeira}}\ and\ \bibinfo {author} {\bibfnamefont {A.~J.}\ \bibnamefont
  {Leggett}},\ }\bibfield  {title} {\bibinfo {title} {Path integral approach to
  quantum brownian motion},\ }\href@noop {} {\bibfield  {journal} {\bibinfo
  {journal} {Physica A: Statistical mechanics and its Applications}\ }\textbf
  {\bibinfo {volume} {121}},\ \bibinfo {pages} {587} (\bibinfo {year}
  {1983}{\natexlab{b}})}\BibitemShut {NoStop}%
\bibitem [{\citenamefont {Verstraten}\ \emph {et~al.}(2022)\citenamefont
  {Verstraten}, \citenamefont {Ludwig}, \citenamefont {Duine},\ and\
  \citenamefont {Smith}}]{verstraten2022}%
  \BibitemOpen
  \bibfield  {author} {\bibinfo {author} {\bibfnamefont {R.~C.}\ \bibnamefont
  {Verstraten}}, \bibinfo {author} {\bibfnamefont {T.}~\bibnamefont {Ludwig}},
  \bibinfo {author} {\bibfnamefont {R.~A.}\ \bibnamefont {Duine}},\ and\
  \bibinfo {author} {\bibfnamefont {C.~M.}\ \bibnamefont {Smith}},\ }\bibfield
  {title} {\bibinfo {title} {The fractional landau-lifshitz-gilbert equation},\
  }\href@noop {} {\bibfield  {journal} {\bibinfo  {journal} {arXiv preprint
  arXiv:2211.12889}\ } (\bibinfo {year} {2022})}\BibitemShut {NoStop}%
\bibitem [{\citenamefont {Kamenev}\ and\ \citenamefont
  {Levchenko}(2009)}]{kamenevandlevchenko2009}%
  \BibitemOpen
  \bibfield  {author} {\bibinfo {author} {\bibfnamefont {A.}~\bibnamefont
  {Kamenev}}\ and\ \bibinfo {author} {\bibfnamefont {A.}~\bibnamefont
  {Levchenko}},\ }\bibfield  {title} {\bibinfo {title} {Keldysh technique and
  non-linear $\sigma$-model: basic principles and applications},\ }\href@noop
  {} {\bibfield  {journal} {\bibinfo  {journal} {Advances in Physics}\ }\textbf
  {\bibinfo {volume} {58}},\ \bibinfo {pages} {197} (\bibinfo {year}
  {2009})}\BibitemShut {NoStop}%
\bibitem [{\citenamefont {Altland}\ and\ \citenamefont
  {Simons}(2010)}]{altlandandsimons2010}%
  \BibitemOpen
  \bibfield  {author} {\bibinfo {author} {\bibfnamefont {A.}~\bibnamefont
  {Altland}}\ and\ \bibinfo {author} {\bibfnamefont {B.~D.}\ \bibnamefont
  {Simons}},\ }\href@noop {} {\emph {\bibinfo {title} {Condensed matter field
  theory}}}\ (\bibinfo  {publisher} {Cambridge university press},\ \bibinfo
  {year} {2010})\BibitemShut {NoStop}%
\bibitem [{\citenamefont {Kamenev}(2011)}]{kamenev2011}%
  \BibitemOpen
  \bibfield  {author} {\bibinfo {author} {\bibfnamefont {A.}~\bibnamefont
  {Kamenev}},\ }\href@noop {} {\emph {\bibinfo {title} {Field theory of
  non-equilibrium systems}}}\ (\bibinfo  {publisher} {Cambridge University
  Press},\ \bibinfo {year} {2011})\BibitemShut {NoStop}%
\bibitem [{\citenamefont {Shnirman}\ \emph {et~al.}(2015)\citenamefont
  {Shnirman}, \citenamefont {Gefen}, \citenamefont {Saha}, \citenamefont
  {Burmistrov}, \citenamefont {Kiselev},\ and\ \citenamefont
  {Altland}}]{shnirman2015}%
  \BibitemOpen
  \bibfield  {author} {\bibinfo {author} {\bibfnamefont {A.}~\bibnamefont
  {Shnirman}}, \bibinfo {author} {\bibfnamefont {Y.}~\bibnamefont {Gefen}},
  \bibinfo {author} {\bibfnamefont {A.}~\bibnamefont {Saha}}, \bibinfo {author}
  {\bibfnamefont {I.~S.}\ \bibnamefont {Burmistrov}}, \bibinfo {author}
  {\bibfnamefont {M.~N.}\ \bibnamefont {Kiselev}},\ and\ \bibinfo {author}
  {\bibfnamefont {A.}~\bibnamefont {Altland}},\ }\bibfield  {title} {\bibinfo
  {title} {Geometric quantum noise of spin},\ }\href@noop {} {\bibfield
  {journal} {\bibinfo  {journal} {Physical Review Letters}\ }\textbf {\bibinfo
  {volume} {114}},\ \bibinfo {pages} {176806} (\bibinfo {year}
  {2015})}\BibitemShut {NoStop}%
\bibitem [{\citenamefont {Ludwig}\ \emph {et~al.}(2019)\citenamefont {Ludwig},
  \citenamefont {Burmistrov}, \citenamefont {Gefen},\ and\ \citenamefont
  {Shnirman}}]{ludwig2019}%
  \BibitemOpen
  \bibfield  {author} {\bibinfo {author} {\bibfnamefont {T.}~\bibnamefont
  {Ludwig}}, \bibinfo {author} {\bibfnamefont {I.~S.}\ \bibnamefont
  {Burmistrov}}, \bibinfo {author} {\bibfnamefont {Y.}~\bibnamefont {Gefen}},\
  and\ \bibinfo {author} {\bibfnamefont {A.}~\bibnamefont {Shnirman}},\
  }\bibfield  {title} {\bibinfo {title} {Thermally driven spin transfer torque
  system far from equilibrium: Enhancement of thermoelectric current via
  pumping current},\ }\href@noop {} {\bibfield  {journal} {\bibinfo  {journal}
  {Physical Review B}\ }\textbf {\bibinfo {volume} {99}},\ \bibinfo {pages}
  {045429} (\bibinfo {year} {2019})}\BibitemShut {NoStop}%
\bibitem [{\citenamefont {Schmid}(1982)}]{schmid1982}%
  \BibitemOpen
  \bibfield  {author} {\bibinfo {author} {\bibfnamefont {A.}~\bibnamefont
  {Schmid}},\ }\bibfield  {title} {\bibinfo {title} {On a quasiclassical
  langevin equation},\ }\href@noop {} {\bibfield  {journal} {\bibinfo
  {journal} {Journal of Low Temperature Physics}\ }\textbf {\bibinfo {volume}
  {49}},\ \bibinfo {pages} {609} (\bibinfo {year} {1982})}\BibitemShut
  {NoStop}%
\bibitem [{\citenamefont {Eckern}\ \emph {et~al.}(1990)\citenamefont {Eckern},
  \citenamefont {Lehr}, \citenamefont {Menzel-Dorwarth}, \citenamefont
  {Pelzer},\ and\ \citenamefont {Schmid}}]{eckern1990}%
  \BibitemOpen
  \bibfield  {author} {\bibinfo {author} {\bibfnamefont {U.}~\bibnamefont
  {Eckern}}, \bibinfo {author} {\bibfnamefont {W.}~\bibnamefont {Lehr}},
  \bibinfo {author} {\bibfnamefont {A.}~\bibnamefont {Menzel-Dorwarth}},
  \bibinfo {author} {\bibfnamefont {F.}~\bibnamefont {Pelzer}},\ and\ \bibinfo
  {author} {\bibfnamefont {A.}~\bibnamefont {Schmid}},\ }\bibfield  {title}
  {\bibinfo {title} {The quasiclassical langevin equation and its application
  to the decay of a metastable state and to quantum fluctuations},\ }\href@noop
  {} {\bibfield  {journal} {\bibinfo  {journal} {Journal of statistical
  physics}\ }\textbf {\bibinfo {volume} {59}},\ \bibinfo {pages} {885}
  (\bibinfo {year} {1990})}\BibitemShut {NoStop}%
\bibitem [{\citenamefont {Weiss}(2012)}]{weiss2012}%
  \BibitemOpen
  \bibfield  {author} {\bibinfo {author} {\bibfnamefont {U.}~\bibnamefont
  {Weiss}},\ }\href@noop {} {\emph {\bibinfo {title} {Quantum dissipative
  systems}}}\ (\bibinfo  {publisher} {World Scientific},\ \bibinfo {year}
  {2012})\BibitemShut {NoStop}%
\bibitem [{Note1()}]{Note1}%
  \BibitemOpen
  \bibinfo {note} {For notational simplicity, since no quantum components
  remain, we drop the index $c$ for classical from now on.}\BibitemShut {Stop}%
\bibitem [{Note2()}]{Note2}%
  \BibitemOpen
  \bibinfo {note} {Note that we have used $\alpha ^A(-\omega ) = \alpha
  ^R(\omega )$. Furthermore, note that an $\omega =0$ contribution as $\alpha
  ^R(0)$ could renormalize the effective magnetic field. In the present case,
  however, the $\omega =0$ contribution is irrelevant for the quasi-classical
  dynamics, as it would lead to a term proportional to $\protect \mathbf S
  \times \protect \mathbf S$, which vanishes identically.}\BibitemShut {Stop}%
\bibitem [{Note3()}]{Note3}%
  \BibitemOpen
  \bibinfo {note} {At low temperatures, the correlation function is given by
  $\langle \xi _m \xi _{m'} \rangle (\omega ) = \alpha _0\protect \, \omega
  \coth (\omega /2k_B T)\protect \, \delta _{mm'}$, where $\omega $ is the
  frequency corresponding to $t-t'$. At high temperatures, $2k_B T \gg \omega
  $, the correlation function simplifies to $\langle \xi _m \xi _{m'} \rangle
  (\omega ) = 2 \alpha _0\protect \, k_B T\protect \, \delta
  _{mm'}$.}\BibitemShut {Stop}%
\bibitem [{\citenamefont {Van~Kampen}(2011)}]{vankampen2011}%
  \BibitemOpen
  \bibfield  {author} {\bibinfo {author} {\bibfnamefont {N.}~\bibnamefont
  {Van~Kampen}},\ }\href@noop {} {\emph {\bibinfo {title} {Stochastic Processes
  in Physics and Chemistry}}}\ (\bibinfo  {publisher} {Elsevier},\ \bibinfo
  {year} {2011})\BibitemShut {NoStop}%
\bibitem [{\citenamefont {R{\"u}ckriegel}\ and\ \citenamefont
  {Kopietz}(2015)}]{ruckriegelandkopietz2015}%
  \BibitemOpen
  \bibfield  {author} {\bibinfo {author} {\bibfnamefont {A.}~\bibnamefont
  {R{\"u}ckriegel}}\ and\ \bibinfo {author} {\bibfnamefont {P.}~\bibnamefont
  {Kopietz}},\ }\bibfield  {title} {\bibinfo {title} {Rayleigh-jeans
  condensation of pumped magnons in thin-film ferromagnets},\ }\href@noop {}
  {\bibfield  {journal} {\bibinfo  {journal} {Physical review letters}\
  }\textbf {\bibinfo {volume} {115}},\ \bibinfo {pages} {157203} (\bibinfo
  {year} {2015})}\BibitemShut {NoStop}%
\bibitem [{Note4()}]{Note4}%
  \BibitemOpen
  \bibinfo {note} {Here, the indices $m,m'$ are defined as above after \protect
  \eqref {eq:iLLG}.}\BibitemShut {Stop}%
\bibitem [{\citenamefont {Streib}\ \emph {et~al.}(2018)\citenamefont {Streib},
  \citenamefont {Keshtgar},\ and\ \citenamefont {Bauer}}]{streib2018damping}%
  \BibitemOpen
  \bibfield  {author} {\bibinfo {author} {\bibfnamefont {S.}~\bibnamefont
  {Streib}}, \bibinfo {author} {\bibfnamefont {H.}~\bibnamefont {Keshtgar}},\
  and\ \bibinfo {author} {\bibfnamefont {G.~E.}\ \bibnamefont {Bauer}},\
  }\bibfield  {title} {\bibinfo {title} {Damping of magnetization dynamics by
  phonon pumping},\ }\href@noop {} {\bibfield  {journal} {\bibinfo  {journal}
  {Physical review letters}\ }\textbf {\bibinfo {volume} {121}},\ \bibinfo
  {pages} {027202} (\bibinfo {year} {2018})}\BibitemShut {NoStop}%
\bibitem [{\citenamefont {Fujii}\ and\ \citenamefont
  {Sakabe}(2001)}]{FUJII20013666}%
  \BibitemOpen
  \bibfield  {author} {\bibinfo {author} {\bibfnamefont {T.}~\bibnamefont
  {Fujii}}\ and\ \bibinfo {author} {\bibfnamefont {Y.}~\bibnamefont {Sakabe}},\
  }\bibfield  {title} {\bibinfo {title} {Growth and magnetic properties of yig
  films},\ }in\ \href
  {https://doi.org/https://doi.org/10.1016/B0-08-043152-6/00654-9} {\emph
  {\bibinfo {booktitle} {Encyclopedia of Materials: Science and Technology}}},\
  \bibinfo {editor} {edited by\ \bibinfo {editor} {\bibfnamefont {K.~J.}\
  \bibnamefont {Buschow}}, \bibinfo {editor} {\bibfnamefont {R.~W.}\
  \bibnamefont {Cahn}}, \bibinfo {editor} {\bibfnamefont {M.~C.}\ \bibnamefont
  {Flemings}}, \bibinfo {editor} {\bibfnamefont {B.}~\bibnamefont {Ilschner}},
  \bibinfo {editor} {\bibfnamefont {E.~J.}\ \bibnamefont {Kramer}}, \bibinfo
  {editor} {\bibfnamefont {S.}~\bibnamefont {Mahajan}},\ and\ \bibinfo {editor}
  {\bibfnamefont {P.}~\bibnamefont {Veyssi{\`e}re}}}\ (\bibinfo  {publisher}
  {Elsevier},\ \bibinfo {address} {Oxford},\ \bibinfo {year} {2001})\ pp.\
  \bibinfo {pages} {3666--3670}\BibitemShut {NoStop}%
\bibitem [{\citenamefont {Kleszczewski}\ and\ \citenamefont
  {Bodzenta}(1988)}]{kleszczewski1988phonon}%
  \BibitemOpen
  \bibfield  {author} {\bibinfo {author} {\bibfnamefont {Z.}~\bibnamefont
  {Kleszczewski}}\ and\ \bibinfo {author} {\bibfnamefont {J.}~\bibnamefont
  {Bodzenta}},\ }\bibfield  {title} {\bibinfo {title} {Phonon--phonon
  interaction in gadolinium--gallium garnet crystals},\ }\href@noop {}
  {\bibfield  {journal} {\bibinfo  {journal} {physica status solidi (b)}\
  }\textbf {\bibinfo {volume} {146}},\ \bibinfo {pages} {467} (\bibinfo {year}
  {1988})}\BibitemShut {NoStop}%
\bibitem [{\citenamefont {Olive}\ \emph {et~al.}(2015)\citenamefont {Olive},
  \citenamefont {Lansac}, \citenamefont {Meyer}, \citenamefont {Hayoun},\ and\
  \citenamefont {Wegrowe}}]{olive2015deviation}%
  \BibitemOpen
  \bibfield  {author} {\bibinfo {author} {\bibfnamefont {E.}~\bibnamefont
  {Olive}}, \bibinfo {author} {\bibfnamefont {Y.}~\bibnamefont {Lansac}},
  \bibinfo {author} {\bibfnamefont {M.}~\bibnamefont {Meyer}}, \bibinfo
  {author} {\bibfnamefont {M.}~\bibnamefont {Hayoun}},\ and\ \bibinfo {author}
  {\bibfnamefont {J.-E.}\ \bibnamefont {Wegrowe}},\ }\bibfield  {title}
  {\bibinfo {title} {Deviation from the landau-lifshitz-gilbert equation in the
  inertial regime of the magnetization},\ }\href@noop {} {\bibfield  {journal}
  {\bibinfo  {journal} {Journal of Applied Physics}\ }\textbf {\bibinfo
  {volume} {117}} (\bibinfo {year} {2015})}\BibitemShut {NoStop}%
\bibitem [{\citenamefont {R{\"u}ckriegel}\ \emph {et~al.}(2014)\citenamefont
  {R{\"u}ckriegel}, \citenamefont {Kopietz}, \citenamefont {Bozhko},
  \citenamefont {Serga},\ and\ \citenamefont
  {Hillebrands}}]{ruckriegel2014magnetoelastic}%
  \BibitemOpen
  \bibfield  {author} {\bibinfo {author} {\bibfnamefont {A.}~\bibnamefont
  {R{\"u}ckriegel}}, \bibinfo {author} {\bibfnamefont {P.}~\bibnamefont
  {Kopietz}}, \bibinfo {author} {\bibfnamefont {D.~A.}\ \bibnamefont {Bozhko}},
  \bibinfo {author} {\bibfnamefont {A.~A.}\ \bibnamefont {Serga}},\ and\
  \bibinfo {author} {\bibfnamefont {B.}~\bibnamefont {Hillebrands}},\
  }\bibfield  {title} {\bibinfo {title} {Magnetoelastic modes and lifetime of
  magnons in thin yttrium iron garnet films},\ }\href@noop {} {\bibfield
  {journal} {\bibinfo  {journal} {Physical Review B}\ }\textbf {\bibinfo
  {volume} {89}},\ \bibinfo {pages} {184413} (\bibinfo {year}
  {2014})}\BibitemShut {NoStop}%
\bibitem [{\citenamefont {Zhukov}\ \emph {et~al.}(2018)\citenamefont {Zhukov},
  \citenamefont {Kirstein}, \citenamefont {Smirnov}, \citenamefont {Yakovlev},
  \citenamefont {Glazov}, \citenamefont {Reuter}, \citenamefont {Wieck},
  \citenamefont {Bayer},\ and\ \citenamefont {Greilich}}]{zhukov2018spin}%
  \BibitemOpen
  \bibfield  {author} {\bibinfo {author} {\bibfnamefont {E.}~\bibnamefont
  {Zhukov}}, \bibinfo {author} {\bibfnamefont {E.}~\bibnamefont {Kirstein}},
  \bibinfo {author} {\bibfnamefont {D.}~\bibnamefont {Smirnov}}, \bibinfo
  {author} {\bibfnamefont {D.}~\bibnamefont {Yakovlev}}, \bibinfo {author}
  {\bibfnamefont {M.}~\bibnamefont {Glazov}}, \bibinfo {author} {\bibfnamefont
  {D.}~\bibnamefont {Reuter}}, \bibinfo {author} {\bibfnamefont
  {A.}~\bibnamefont {Wieck}}, \bibinfo {author} {\bibfnamefont
  {M.}~\bibnamefont {Bayer}},\ and\ \bibinfo {author} {\bibfnamefont
  {A.}~\bibnamefont {Greilich}},\ }\bibfield  {title} {\bibinfo {title} {Spin
  inertia of resident and photoexcited carriers in singly charged quantum
  dots},\ }\href@noop {} {\bibfield  {journal} {\bibinfo  {journal} {Physical
  Review B}\ }\textbf {\bibinfo {volume} {98}},\ \bibinfo {pages} {121304}
  (\bibinfo {year} {2018})}\BibitemShut {NoStop}%
\bibitem [{\citenamefont {Smirnov}\ \emph {et~al.}(2018)\citenamefont
  {Smirnov}, \citenamefont {Zhukov}, \citenamefont {Kirstein}, \citenamefont
  {Yakovlev}, \citenamefont {Reuter}, \citenamefont {Wieck}, \citenamefont
  {Bayer}, \citenamefont {Greilich},\ and\ \citenamefont
  {Glazov}}]{smirnov2018theory}%
  \BibitemOpen
  \bibfield  {author} {\bibinfo {author} {\bibfnamefont {D.}~\bibnamefont
  {Smirnov}}, \bibinfo {author} {\bibfnamefont {E.}~\bibnamefont {Zhukov}},
  \bibinfo {author} {\bibfnamefont {E.}~\bibnamefont {Kirstein}}, \bibinfo
  {author} {\bibfnamefont {D.}~\bibnamefont {Yakovlev}}, \bibinfo {author}
  {\bibfnamefont {D.}~\bibnamefont {Reuter}}, \bibinfo {author} {\bibfnamefont
  {A.}~\bibnamefont {Wieck}}, \bibinfo {author} {\bibfnamefont
  {M.}~\bibnamefont {Bayer}}, \bibinfo {author} {\bibfnamefont
  {A.}~\bibnamefont {Greilich}},\ and\ \bibinfo {author} {\bibfnamefont
  {M.}~\bibnamefont {Glazov}},\ }\bibfield  {title} {\bibinfo {title} {Theory
  of spin inertia in singly charged quantum dots},\ }\href@noop {} {\bibfield
  {journal} {\bibinfo  {journal} {Physical Review B}\ }\textbf {\bibinfo
  {volume} {98}},\ \bibinfo {pages} {125306} (\bibinfo {year}
  {2018})}\BibitemShut {NoStop}%
\end{thebibliography}%

%\foreach \x in {1,...,4} 
%{% 
%\clearpage 
%\includepdf[pages={\x},turn=false]{Supplement-for-arxiv.pdf}
%}

%\clearpage
%\includepdf[pages={1,2,3,4},turn=false]{supplement.pdf}



\section*{Supplementary material}

# SUPPLEMENTAL MATERIAL: BATH-INDUCED SPIN INERTIA

## A SIMPLE CHOICE OF GAUGE IS SUFFICIENT

Before choosing a gauge, the action, Eq. (1) in the main text, still contains the gauge freedom $\psi$ in the Berry-phase term, which becomes $-i\langle \dot{g}|g\rangle = S\,(\dot\psi + \dot\phi\cos\theta)$ in the Euler-angle representation. In principle, we can fix $\psi$ as we like. However, we should respect "boundary conditions" on the Keldysh contour; otherwise, we could choose $\dot\psi = -\dot\phi\cos\theta$, which would eliminate the Berry-phase term $\oint_K dt\, S\,(\dot\psi + \dot\phi\cos\theta)$ from the action. It can be crucial to be careful with the boundary conditions of the chosen gauge [1, 2]. Here, however, a simple choice of $\dot\psi = -\dot\phi$ on the Keldysh contour will be just fine; as shown next, it leads to the same quasi-classical dynamics as the more elaborate choice of references [1, 2].

To compare the gauge choice of references [1, 2] with the simple choice $\dot\psi = -\dot\phi$, we rewrite $\psi = \chi - \phi$. Then, the gauge choices differ only in $\chi$; references [1, 2] choose $\dot\chi_c = \dot\phi_c(1 - \cos\theta_c)$ and $\chi_q = \phi_q(1 - \cos\theta_c)$, while the simple choice corresponds to $\chi = 0$. For the Berry-phase term, there is no relevant difference at finite times:

$$\oint_K dt\,(\dot\psi + \dot\phi\cos\theta) = -\oint_K dt\,\dot\phi\,(1-\cos\theta) + \oint_K dt\,\dot\chi\;; \tag{1}$$

but with $\chi_\pm = \chi_c \pm \chi_q/2$ and $\chi_q = \phi_q(1-\cos\theta_c)$, we find

$$\oint_K dt\,\dot\chi = \int dt\,[\dot\phi_q(1-\cos\theta_c) + \phi_q\dot\theta_c\sin\theta_c] = 0\;, \tag{2}$$

where, in the last step, we used integration by parts and disregarded potentially arising boundary terms that are dynamically irrelevant in the quasiclassical approximation.

## DERIVATION OF BATH SPECTRAL DENSITY FOR A PHONON BATH

Spins on a lattice will be coupled to the phonons that arise from oscillations of the atoms/ions at the lattice sites. In spintronics, this coupling is often described by the magnetoelastic coupling. For example, see [3], where they derived the coupled equations of motions for spins $\mathbf{S}_i$ on a lattice (with lattice sites denoted by $i$) that are coupled to a phonon bath,

$$\dot{\mathbf{S}}_i = \mathbf{S}_i \times \mathbf{H} + \mathbf{S}_i \times \mathbf{h}_i + \mathbf{S}_i \times \sum_j K_{ij}\mathbf{S}_j - \mathbf{S}_i \times \int_0^t dt' \sum_j G_{ij}(t,t')\dot{\mathbf{S}}_j(t')\;, \tag{3}$$

where $\mathbf{H}$ is the external magnetic field, $K_{ij}$ accounts for dipolar and nearest-neighbor exchange interactions, and the coupling to the phonon system gives rise to the induced magnetic field $\mathbf{h}_i$ and the generalized Gilbert damping with the kernel function $G_{ij}(t,t')$. For the components of that kernel function, they find

$$G_{ij}^{mm'}(t,t') = \frac{1}{NS_1^4}\sum_{\tilde m \tilde m'} B_{m\tilde m}B_{m'\tilde m'}S_i^{\tilde m}(t)S_j^{\tilde m'}(t')\sum_{\mathbf{k}\lambda} e^{i\mathbf{k}\cdot(\mathbf{R}_i - \mathbf{R}_j)}(\mathbf{k}_{m\tilde m}\cdot\mathbf{e}_{\mathbf{k}\lambda})(\mathbf{k}_{m'\tilde m'}\cdot\mathbf{e}_{-\mathbf{k}\lambda})\frac{\cos[\omega_{\mathbf{k}\lambda}(t-t')]}{M\omega_{\mathbf{k}\lambda}^2}\;, \tag{4}$$

where $S_1$ is the spin length of the effective spin on a single lattice site with spin, $N$ is the number of lattice sites, $\mathbf{k}$ denotes the phonon momentum, $\lambda$ its polarization (longitudinal/transversal), and the corresponding dispersion relation and polarization vector are given by $\omega_{\mathbf{k}\lambda}$ and $\mathbf{e}_{\mathbf{k}\lambda}$ respectively. Further, $N$ is the number of lattice sites, $M$ is their effective ionic mass, and $B_{m\tilde m}$ are the magnetoelastic coupling coefficients. We also adopt their short notation for $\mathbf{k}_{m\tilde m} = k_m\mathbf{e}_{\tilde m} + k_{\tilde m}\mathbf{e}_m$ with Cartesian unit vectors $\mathbf{e}_m$ for $m \in x, y, z$.

Now, to find the bath spectral density for the phonon bath, our intermediate goal is to relate the spin-lattice equation of motion (3) to the macrospin equation of motion of the main text (2); more specifically, we need to relate the generalized Gilbert-damping terms of both equations. This immediately leads us to a key problem: the kernel function for the phonon bath, $G_{ij}^{mm'}(t,t')$, depends on the lattice spins via $S_i^{\tilde m}(t)S_j^{\tilde m'}(t')$, whereas the kernel function for the harmonic-oscillator bath of the main text, $\tilde\alpha(t-t')$, does not depend on the macrospin. This key difference arises from the different spin-to-bath couplings. While the magnetoelastic coupling is nonlinear in the spin [3], in the main text we used the Caldeira-Leggett approach with a simple linear spin-to-bath coupling. However, if the spins

remain close to the $z$-direction, such that we can linearize the equation of motion in deviations from the $z$-direction, then we may simply approximate $\mathbf{S}_i \approx S_1 \mathbf{e}_z$ in $G_{ij}^{mm'}(t,t')$. In turn, the kernel function simplifies to

$$G_{ij}^{mm'}(t-t') = \frac{1}{NS_1^2} B_{mz} B_{m'z} \sum_{\mathbf{k}\lambda} e^{i\mathbf{k}\cdot(\mathbf{R}_i-\mathbf{R}_j)} (\mathbf{k}_{mz}\cdot\mathbf{e}_{\mathbf{k}\lambda})(\mathbf{k}_{m'z}\cdot\mathbf{e}_{-\mathbf{k}\lambda}) \frac{\cos[\omega_{\mathbf{k}\lambda}(t-t')]}{M\omega_{\mathbf{k}\lambda}^2} \; . \tag{5}$$

With this simplification, we can now Fourier-transform equation (3) with $\sum_i e^{-i\mathbf{q}\mathbf{R}_i}...$ and use the macrospin approximation with $\mathbf{S}_{\mathbf{q}=0} = \mathbf{S}$ and $\mathbf{S}_{\mathbf{q}\neq 0} = 0$. For the generalized Gilbert damping, we then find

$$-\mathbf{S}_i \times \int_0^t dt' \sum_j G_{ij}(t,t') \dot{\mathbf{S}}_j(t') \quad \longrightarrow \quad -\frac{1}{N_s} \mathbf{S} \times \int_{-\infty}^t dt' \, G_0(t-t') \dot{\mathbf{S}}(t') \; , \tag{6}$$

where $N_s$ is the number of lattice sites with spin and, disregarding transient effects, we set the lower integral boundary from 0 to $-\infty$. The kernel function $G_0(t-t')$ is the $\mathbf{q}=0$ case of the Fourier-transform

$$G_{\mathbf{q}}(t-t') = \sum_i e^{i\mathbf{q}(\mathbf{R}_i-\mathbf{R}_j)} G_{ij}(t-t') \; . \tag{7}$$

Now, the generalized Gilbert damping, Eq. (6), can be integrated by parts to find

$$-\frac{1}{N_s} \mathbf{S} \times \int_{-\infty}^t dt' \, G_0(t-t') \dot{\mathbf{S}}(t') = -\frac{1}{N_s} \mathbf{S} \times \left[ \int_{-\infty}^t dt' \, \dot{G}_0(t-t') \mathbf{S}(t') + [G_0(t-t')\mathbf{S}(t')]_{-\infty}^t \right] \; , \tag{8}$$

where the boundary term $[G_0(t-t')\mathbf{S}(t')]_{-\infty}^t$ only contains a term that renormalizes the magnetic field and some information about the initial state, which would only affect transient dynamics that we are not considering anyways.

So, we can finally relate the generalized Gilbert damping for the phonon bath to the generalized Gilbert damping for the Caldeira-Leggett approach in the main text. We find that the relation between the two kernel functions is given by

$$\tilde{\alpha}(t-t') = -\frac{1}{N_s} \dot{G}_0(t-t') \Theta(t-t') \; . \tag{9}$$

Subtracting the zero-frequency part $\tilde{\alpha}(\omega = 0)$ for regularization, as in the main text, and using the relation

$$\int_{-\infty}^\infty \frac{d\omega}{2\pi} e^{-i\omega t} \frac{1}{(\omega+i\eta)^2 - \epsilon^2} = -\frac{\sin(\epsilon\, t)}{\epsilon} \Theta(t) \; , \tag{10}$$

we find the (tensorial) bath spectral density for the phonon bath,

$$J_{mm'}(\epsilon) = \pi \sum_i \sum_{\mathbf{k}\lambda} e^{i\mathbf{k}\mathbf{R}_i} \frac{B_{mz} B_{m'z}}{2MNS_1 S\, \omega_{\mathbf{k}\lambda}} (\mathbf{k}_{mz}\cdot\mathbf{e}_{\mathbf{k}\lambda})(\mathbf{k}_{m'z}\cdot\mathbf{e}_{-\mathbf{k}\lambda}) \delta(\epsilon - \omega_{\mathbf{k}\lambda}) \; , \tag{11}$$

where we used that the macrospin length $S$ is given by the sum over all the individual spin lengths $S_1$; so, $S = N_s S_1$. The resulting bath spectral density, Eq. (11), is used in the main text.

### EXPLICIT CALCULATION FOR A PLANAR SIMPLE-CUBIC SPIN LATTICE IN A THREE-DIMENSIONAL SIMPLE-CUBIC LATTICE

We consider a three-dimensional simple cubic lattice with site length $a$, where the $z=0$-plane (YIG) contains spins of length $S_1$ on every site, while there is no spin on all the other sites (GGG).

Starting from the bath spectral density, Eq. (11), we can now use that the sum over lattice sites with spin yields $\sum_i \exp[i\mathbf{k}\mathbf{R}_i] = \sum_{n_x n_y} \exp[i(k_x a\, n_x + k_y a\, n_y)] = N_x \delta_{k_x,0} N_y \delta_{k_y,0}$, where $n_x$ and $n_y$ numerate, respectively, the lattice sites in $x$-direction and $y$-direction; and $N_x$ and $N_y$ are the number of lattice sites in $x$-direction and $y$-direction respectively. Consequently, the bath spectral density becomes

$$J_{mm'}(\epsilon) = \pi \sum_{\mathbf{k}\lambda} N_x \delta_{k_x,0} N_y \delta_{k_y,0} \frac{B_{mz} B_{m'z}}{2MNS_1 S\, \omega_{\mathbf{k}\lambda}} (\mathbf{k}_{mz}\cdot\mathbf{e}_{\mathbf{k}\lambda})(\mathbf{k}_{m'z}\cdot\mathbf{e}_{-\mathbf{k}\lambda}) \delta(\epsilon - \omega_{\mathbf{k}\lambda}) \; , \tag{12}$$

and, after carrying out the sums over $k_x$ and $k_y$, it simplifies to

$$J_{mm'}(\epsilon) = \delta_{mm'} \frac{\pi(1+\delta_{mz})^2 B_{mz}^2}{2MS_1 S} \sum_{k_z \lambda} \frac{k_z^2}{N_z \,\omega_{k_z \lambda}} \delta_{m\lambda}\, \delta(\epsilon - \omega_{k_z \lambda}) \;, \tag{13}$$

where we used that the number of lattice sites is given by $N = N_x N_y N_z$ with the number of lattice sites in $z$-direction $N_z$. Carrying out the summation over the phonon polarization yields,

$$J_{mm'}(\epsilon) = \delta_{mm'} \frac{\pi(1+\delta_{mz})^2 B_{mz}^2}{2MS_1 S} \sum_{k_z} \frac{k_z^2}{N_z \,\omega_{k_z m}} \delta(\epsilon - \omega_{k_z m}) \;. \tag{14}$$

Instead of directly carrying out the sum over $k_z$, which runs over all $k_z$-states in the Brillouin zone (BZ), we switch to the continuum case and turn it into an integral; namely, we replace $\sum_{k_z \in BZ} \ldots = (L_z/2\pi) \int_{-\pi/a}^{\pi/a} dk_z \ldots$, where $L_z$ is the length of the system in $z$-direction. Further, using that $L_z = a N_z$, we find

$$J_{mm'}(\epsilon) = \delta_{mm'} \frac{(1+\delta_{mz})^2 B_{mz}^2 a}{4MS_1 S} \int_{-\pi/a}^{\pi/a} dk_z \, \frac{k_z^2}{\omega_{k_z m}} \delta(\epsilon - \omega_{k_z m}) \;. \tag{15}$$

To carry out the remaining integral, we need to know the dispersion relation. For a simple cubic lattice, focusing on acoustic phonons, the dispersion relation is given by $\omega_{k_z m} = (2/a) v_m |\sin(k_z a/2)|$, where $v_m$ is the sound velocity for transversal ($m \in \{x,y\}$) or longitudinal ($m = z$) polarization; see [4, 5] for example. In terms of microscopic parameters, the sound velocities are given by $v_m = \sqrt{K_m/M}$ with the force constants $K_x = K_y = K_2$ and $K_z = K_1 + 2K_2$, where $K_1$ and $K_2$ are respectively the force constants for nearest neighbor and next-nearest neighbor restoring forces [4, 5]. Now, the momentum integral becomes

$$\int_{-\pi/a}^{\pi/a} dk_z \, \frac{k_z^2}{\omega_{k_z m}} \delta(\epsilon - \omega_{k_z m}) = \int_{-\pi/a}^{\pi/a} dk_z \, \frac{k_z^2}{\frac{2v_m}{a} |\sin(\frac{k_z a}{2})|} \, \delta\!\left(\epsilon - \frac{2v_m}{a}|\sin(\frac{k_z a}{2})|\right) = \begin{cases} 2\dfrac{\frac{4}{a^2}\arcsin^2 \frac{a\epsilon}{2v_m}}{v_m \epsilon \sqrt{1-\left(\frac{a\epsilon}{2v_m}\right)^2}} & \text{for } \epsilon \in [0, \frac{2v_m}{a}] \\ 0 & \text{otherwise} \end{cases}. \tag{16}$$

In turn, we find the bath spectral density in explicit form

$$J_{mm'}(\epsilon) = \delta_{mm'} \frac{(1+\delta_{mz})^2 B_{mz}^2}{MS_1 S} \frac{\frac{4}{a^2}\arcsin^2 \frac{a\epsilon}{2v_m}}{\frac{2v_m}{a}\epsilon \sqrt{1-\left(\frac{a\epsilon}{2v_m}\right)^2}} \,\Theta(\epsilon)\, \Theta\!\left(\frac{2v_m}{a} - \epsilon\right) \;. \tag{17}$$

At low energies, $\epsilon \ll \frac{2v_m}{a}$, we can approximate $\arcsin\!\left(\frac{a\epsilon}{2v_m}\right) \approx \frac{a\epsilon}{2v_m}$ and $\sqrt{1-\left(\frac{a\epsilon}{2v_m}\right)^2} \approx 1$, such that we find the low-energy or low-frequency approximation for the bath spectral density

$$J_{\text{lf},mm'}(\epsilon) = \delta_{mm'} \frac{(1+\delta_{mz})^2 B_{mz}^2 a}{2MS_1 S\, v_m^3} \,\epsilon\, \Theta(\epsilon), \tag{18}$$

which is linear in $\epsilon$. So, following the derivation of the main text, we can simply read off the (tensorial) Gilbert-damping coefficient as

$$\alpha_{mm'} = \delta_{mm'} \frac{(1+\delta_{mz})^2 B_{mz}^2 a}{2MS_1 S\, v_m^3} \;. \tag{19}$$

Knowing the bath spectral density $J_{mm'}(\epsilon)$ and its low-frequency approximation $J_{\text{lf},mm'}(\epsilon)$, we can now determine the bath-induced spin inertia from Eq. (7) from the main text,

$$I_{mm'} = \frac{2}{\pi} \int_{-\infty}^{\infty} d\epsilon \, \frac{J_{mm'}(\epsilon) - J_{\text{lf},mm'}(\epsilon)}{\epsilon^3} \;. \tag{20}$$

Inserting Eqs. (17) and (18), and using Eq. (19), we find

$$I_{mm'} = \frac{2}{\pi} \alpha_{mm'} \int_0^{\infty} d\epsilon \left[ \frac{4v_m^2}{a^2} \frac{\arcsin^2 \frac{a\epsilon}{2v_m}}{\epsilon^4 \sqrt{1-\left(\frac{a\epsilon}{2v_m}\right)^2}} \Theta\!\left(\frac{2v_m}{a} - \epsilon\right) - \frac{1}{\epsilon^2} \right] \;. \tag{21}$$

The $\epsilon$-integral can be made dimensionless by substituting $\epsilon = 2v_m\epsilon'/a$, which yields

$$I_{mm'} = \frac{1}{\pi}\frac{a}{v_m}\alpha_{mm'}\int_0^\infty d\epsilon' \left[\frac{\arcsin^2\epsilon'}{\epsilon'^4\sqrt{1-\epsilon'^2}}\Theta(\tfrac{2v_m}{a}-\tfrac{2v_m}{a}\epsilon') - \frac{1}{\epsilon'^2}\right]. \qquad (22)$$

To deal with the $\Theta$-function, it is now convenient to split the integral from 0 to $\infty$ into two parts: one from 0 to 1 and another one from 1 to $\infty$. Then, we find

$$\int_0^\infty d\epsilon'\left[\frac{\arcsin^2\epsilon'}{\epsilon'^4\sqrt{1-\epsilon'^2}}\Theta(\tfrac{2v_m}{a}-\tfrac{2v_m}{a}\epsilon') - \frac{1}{\epsilon'^2}\right] = \underbrace{\int_0^1 d\epsilon'\left[\frac{\arcsin^2\epsilon'}{\epsilon'^4\sqrt{1-\epsilon'^2}} - \frac{1}{\epsilon'^2}\right]}_{\approx 1.9} + \underbrace{\int_1^\infty d\epsilon'\left[-\frac{1}{\epsilon'^2}\right]}_{=-1} \approx 0.9, \qquad (23)$$

where we evaluated the integral from 0 to 1 numerically. So, overall, for a layer of YIG sandwiched between two bulk layers of GGG, we find the phonon-bath-induced spin inertia

$$I_{mm'} = \frac{0.9}{\pi}\frac{a}{v_m}\alpha_{mm'}, \qquad (24)$$

as given in the main text.

---

[1] A. Shnirman, Y. Gefen, A. Saha, I. S. Burmistrov, M. N. Kiselev, and A. Altland, Geometric quantum noise of spin, Physical Review Letters **114**, 176806 (2015).
[2] T. Ludwig, I. S. Burmistrov, Y. Gefen, and A. Shnirman, Thermally driven spin transfer torque system far from equilibrium: Enhancement of thermoelectric current via pumping current, Physical Review B **99**, 045429 (2019).
[3] A. Rückriegel and P. Kopietz, Rayleigh-jeans condensation of pumped magnons in thin-film ferromagnets, Physical review letters **115**, 157203 (2015).
[4] O. Madelung, *Introduction to solid-state theory*, Vol. 2 (Springer Science & Business Media, 1996).
[5] J. Sólyom, *Fundamentals of the Physics of Solids* (Springer, 2007).



\end{document}